\documentclass[lettersize,journal]{IEEEtran}

\ifCLASSOPTIONcompsoc
  \usepackage[nocompress]{cite}
\else
  \usepackage{cite}
\fi
\usepackage{amssymb,amsmath,amsthm}

\usepackage{blindtext, graphicx}

\usepackage{subcaption}

\usepackage[utf8]{inputenc}

\usepackage{lineno}

\usepackage{colortbl}

\usepackage{pifont}

\usepackage{verbatim}

\usepackage{balance}

\usepackage[caption=false]{subfig}

\usepackage{cancel}

\usepackage[color=pink]{todonotes}

\newcommand{\olga}[1]{\textcolor{purple}{(Olga: #1)}}

\usepackage{algorithm}
\usepackage[noend]{algpseudocode} 
\algrenewcommand\alglinenumber[1]{\tiny #1:}
\algdef{SE}[SUBALG]{Indent}{EndIndent}{}{\algorithmicend\ }%
\algtext*{Indent}
\algtext*{EndIndent}

\usepackage{url}
\usepackage{tikz}
\usetikzlibrary{shapes,arrows}
\usetikzlibrary{arrows.meta, shapes.geometric, calc, shadows,fit,matrix,positioning}
\usetikzlibrary{patterns}
\tikzset{
	>=stealth',
	punkt/.style={
		rectangle,
		rounded corners,
		draw=black, thick,
		text width=8em,
		minimum height=2em,
		text centered},
	legend/.style={
		rectangle,
		rounded corners,
		draw=white, very thick,
		text width=40em,
		minimum height=1.7em},
	pil/.style={
		->,
		thick,
		shorten <=1.5pt,
		shorten >=1.5pt,}
}

\tikzset
{
	treenode/.style = {circle, draw=black, align=center, minimum size=0.7cm},
	requesternode/.style = {circle, draw=black, align=center, minimum size=0.7cm,line width=0.3mm},
	masternode/.style = {circle,densely dotted, draw=black, align=center, minimum size=0.7cm,line width=0.3mm},
	subtree/.style  = {isosceles triangle, draw=black, align=center, minimum height=0.3cm, minimum width=0.5cm,shape border rotate=90, anchor=north}
}

\begin{document}

\title{Opportunistic Optical Link Scheduling in the \\ Scalable Matching Model}

\author{Marcos~T.~Henriques~and~Olga~Goussevskaia
\IEEEcompsocitemizethanks{\IEEEcompsocthanksitem M.T. Henriques and O. Goussevskaia are with the Computer Science Department, Universidade Federal de Minas Gerais (UFMG), 31270-901 Belo Horizonte, Brazil.\protect\\
E-mails: marcoshenriques@dcc.ufmg.br, olga@dcc.ufmg.br
}
}

\maketitle

\begin{abstract}
The Scalable Matching Model (SMM) is a useful abstraction for reconfigurable datacenter networks. In SMM, a set of $n$ (Top-of-Rack) electrical packet switches is connected via a set of optical circuit switches (OCS), each realizing a $p \times p$ matching, where $p$ is the number of ports per switch. In the case when $p \ll n$, bounded-degree networks utilize only a small portion of the OCS infrastructure at each point in time, especially when traffic patterns are skewed and bursty.

In this work, we propose the Opportunistic Optical Link Scheduling (OOLS), a set of lightweight OCS link scheduling routines, designed to augment bounded-degree reconfigurable network topologies by leveraging idle OCS resources. We evaluate the performance of OOLS in the SMM through simulations, using synthetic and real-world data-center workloads. Our simulation results show that OOLS substantially improves performance, lowering communication costs and boosting throughput.
\end{abstract}

\begin{IEEEkeywords}
Reconfigurable Data Center Networks, Self-Adjusting Networks, Optical Circuit Switching, Heuristics
\end{IEEEkeywords}

\IEEEdisplaynontitleabstractindextext
\IEEEpeerreviewmaketitle

\section{Introduction}

The growing reliance on data-centric applications across various domains, coupled with advances in AI and machine learning, has significantly increased communication traffic in data center networks \cite{2023WWW_traffic, 2022Jupiter}. Traditionally, data center network designs have relied on static topologies, like Clos \cite{singh2015jupiter, f10}, hypercube \cite{bcube, mdcube}, or expander \cite{xpander, jellyfish}. Such architectures are essentially demand-oblivious and perform well for uniform all-to-all traffic patterns. Traffic in data centers, however, is known to be bursty and skewed \cite{2023WWW_traffic, trace-complexity}.

Recent progress in optical circuit switching (OCS) has enabled demand-aware reconfigurable data center networks (RDNs) to emerge \cite{2025CACM, 2024RDCN, 2021SurveyRONs, 2024NegotiaToR, sigmetrics23mars, sigmetrics23duo}. Several technologies support this new paradigm, including nanosecond-scale tunable lasers \cite{sirius}, digital micromirror devices \cite{projector}, liquid-crystal-on-silicon wavelength-selective switches \cite{2017WSS}, and microcombs and
semiconductor optical amplifiers \cite{2021nature}.

One common abstraction for RDNs is the ToR-Matching-ToR (TMT) model \cite{infocom22lazy, spaa21lazy, avin2020online, infocom23matching, Hanauer_2022}, in which $n$ (Top-of-Rack) electrical \textit{packet} switches are linked via a set of optical \textit{circuit} switches, each realizing a $p \times p$ matching, where $p$ is the number of OCS ports per switch. 
Interesting optimization problems arise in this context, e.g., regarding which is the best combination between electrical packet switches and OCSs \cite{2023TNSM, 2025TNSM3}. 
There is still, however, no consensus in the community on which RDN design is optimal \cite{2025CACM}.


TMT can be used as a testbed for
Self-Adjusting Networks (SAN) \cite{ccr18san, 2025CACM, apocs21renets, splaynet, 2021TCCDisplaynet, displaynet, 2021CBNet, 2022COMNETCBNet}. SANs optimize their physical topology toward the demand they serve in an online manner, i.e., without prior knowledge of the traffic, 
with the requirement that the network remain connected and have bounded degree \cite{disc17}.
Previous studies have shown that, in the case when $p \ll n$, SANs typically utilize only a small portion of the OCS infrastructure at each point in time, especially when spatial and temporal locality of the workload is high, i.e. traffic is skewed and bursty \cite{2023infocomOpticnet, 2024tccOpticnet}. This unexplored idle capacity motivates us to develop opportunistic scheduling strategies for these network architectures. 

Our work is based on the Scalable Matching Model (SMM) \cite{2024tccOpticnet, 2023infocomOpticnet}, a generalization of the TMT model \cite{avin2020online}. SMM introduces the constraint that the number of ports of each OCS is bounded by a constant, i.e., $p=O(1)$. SMM guarantees strict non-blocking (i.e., any point-to-point connection can always be established), uses a minimum number of OCSs to realize any dynamic bounded-degree topology, and allows to execute existing SAN algorithms on top of it, while preserving the performance (amortized communication cost) guarantees of the latter.


We present the Opportunistic Optical Link Scheduling (OOLS), a set of lightweight OCS link scheduling routines, designed to integrate with TMT RDN architectures. OOLS leverages idle OCS ports to dynamically augment a SAN topology with opportunistic routing shortcuts. We implemented OOLS in combination with multiple SAN algorithms using OpticNet \cite{2024tccOpticnet}, a framework for simulating RDNs in the TMT model. 
We performed extensive simulations using both synthetic and real data center workloads, with variable temporal and spatial locality. 
The simulation results show that OOLS significantly increases the \textit{concurrency} of the RDN system and
optimizes not only the system's throughput and makespan,
but also the total communication cost.
Our experiments also revealed a trade-off between efficiency improvement and resource utilization, as a function of $p$ (the number of OCS ports per switch).
In particular, the smaller the OCS, the higher the potential gains of opportunistic link scheduling. Overall, OOLS proved to be an effective and efficient mechanism to utilize idle OCS capacity in RDNs, especially in network architectures that utilize OCSs of limited size.

In Sec. \ref{sec:model}, we present the model and necessary notation; in Sec. \ref{sec:ools}, the OOLS algorithm; in Sec. \ref{sec:evaluation}, the evaluation framework; in Sec. \ref{sec:results}, the simulation results; in Sec. \ref{sec:related}, the related work; finally, in Sec. \ref{sec:conclusion}, we present our conclusion.


\section{Model}
\label{sec:model}

We consider a network topology that connects a set $V$ of $n$ communication nodes (e.g., top-of-rack switches, ToRs), and a sequence $\sigma$ of $m$ messages $\sigma_i(s_i, d_i)\in V \times V$, with source~$s_i$ and destination~$d_i$; $m$ can be infinite. The sequence~$\sigma$ is revealed over time in an online manner.

\textbf{SANs:} SANs are designed to minimize the communication cost through a sequence of network topologies ~$N_0, N_1, \ldots$, where each~$N_t=(V, E_t)$ belongs to some desired graph family $\mathcal{N}$. For scalability reasons, the networks should have \textit{bounded degree} ($\leq k$).

\textbf{Adjustments:} To adjust the topology \textit{smoothly} over time, the network is reconfigured (locally) through adjustments that preserve the desired properties of the graph family $\mathcal{N}$. SANs like SplayNet \cite{splaynet} are based on Binary Search Trees (BST) and extend the classical \textit{zig, zig-zig} and \textit{zig-zag} rotations, first introduced for splay trees \cite{splaytrees}. 
Each splay operation updates a constant number of links, at a constant cost, and roughly halves the depth of every node along the communication path. This halving effect makes splaying efficient in an \textit{amortized cost} analysis.

\textbf{TMT:} The TMT \cite{spaa21lazy, avin2020online} is a model for reconfigurable data center networks (RDN), based on a two-layer architecture, in which a set of (leaf-layer) ToRs is connected using a set of (spine-layer) reconfigurable Optical Circuit Switches (OCS). Each OCS dynamically connects its input-output port-pairs via a \textit{matching}\footnote{A matching is defined as a set of at most $p$ non-adjacent input–output port pairs.}, which changes over time.

\textbf{SMM:} The Scalable Matching Model \cite{2023infocomOpticnet, 2024tccOpticnet} is a generalization of the TMT model \cite{spaa21lazy, avin2020online}. 
Let $\mathcal{N}$ be the family of bounded degree ($\leq k$) graphs on a set $V$ of $n$ nodes (static leaf-layer ToR switches) and $\mathcal{N}_t\in\mathcal{N}$ be a network topology at time $t$. Let $SW = \{sw_1, \ldots, sw_{n'}\}$ be a set of OCSs, of size $n' = f(n,p,k)$, where each switch $sw_i\in SW$ internally connects its $p=\mathcal{O}(1)$ input-output port pairs via a matching $M(i,t)$ of size $\leq p$. At each time $t$, the network $N_t$ is equal to the union of all these matchings: $N_t = \mathcal{M}_t = \cup_{i=1}^{n'} M(i,t)$.

\textbf{TMT port configuration:} Before the RDN starts operating, the TMT architecture must be configured, such that for each (undirected) edge in the network topology $N_t$, there must be a static (duplex) link connecting an output port of $u$ to an input port of some OCS $sw_x \in SW$, i.e., $\exists sw_x \in SW | (u,v)\in M(x,t), \forall (u,v)\in N_t$.

\textbf{Time model:} Time is divided into \textit{synchronous} rounds, and each round consists of a constant number of time slots. In a round, multiple (independent) nodes can make adjustments \textit{concurrently}. We assume that nodes and communication between them are reliable. 

\textbf{Makespan:} Consider a sequence $\sigma$ of $m$ messages, a SAN algorithm $\mathcal{A}$, any initial network $N_{0}\in \mathcal{N}$, and a message $\sigma_i(s_i, d_i)\in\sigma$, generated at time $b_i$ and delivered at time $e_i$.
In terms of time, the objective of a SAN is to minimize the \textit{makespan}, defined as: $\max\limits_{1 \leq i \leq m}{e_i} - b_i$. \label{eq:makespan}

\textbf{Work:} The communication cost can be partitioned into reconfiguration and routing costs. The reconfiguration cost is proportional to the number of (link) adjustments performed by a SAN algorithm $\mathcal{A}$ to deliver a message $\sigma_i\in\sigma$. If $\alpha$ is the cost of a single edge adjustment, then the \textbf{reconfiguration cost} for the entire network is defined as
$linkAdj(N_{t-1}, N_t) = \alpha\cdot \sum_{sw_x\in SW} |M(x,t) \setminus M(x,t-1)|$. The service, or \textbf{routing cost} of a message $\sigma_i\in\sigma$ is proportional to the length of the (shortest) path $dist_{N_{e_i}}(s_i,d_i), N_{e_i}\in\mathcal{N}$. We assume that routing a message incurs a cost of $1$ unit per hop ($(u,v)\in N_t$).

Consider any initial network~$N_{0}$, a sequence of $m$ messages~$\sigma_i(s_i, d_i)\in\sigma$ and algorithm $\mathcal{A}$. Let $t_{\max}=\max\limits_{1 \leq i \leq m}{e_i}$. We define the \textbf{total communication cost} as:
\begin{eqnarray}
\mathcal{C}(\mathcal{A}, N_0, \sigma) &=& \mathcal{D}(\mathcal{A}, N_0, \sigma) +\mathcal{R}(\mathcal{A}, N_0, \sigma), ~\text{where} \label{eq:total_cost}\\
\mathcal{D}(\mathcal{A}, N_0, \sigma) &=& \sum_{i=1}^{m} (dist_{N_{e_i}}(s_i,d_i) + 1),\label{eq:srv_cost}\\ 
\mathcal{R}(\mathcal{A}, N_0, \sigma) &=& \alpha \cdot \sum_{t=1}^{t_{\max}}{linkAdj(N_{t-1}, N_t)}. \label{eq:link_cost}
\end{eqnarray}

\textbf{OpticNet RDN:}
OpticNet \cite{2023infocomOpticnet, 2024tccOpticnet} is a framework for modeling TMT architectures and simulating SAN algorithms in RDN environments. OpticNet partitions the set of nodes  $V$ into $C = \left\lceil {n}/{p}\right\rceil$ clusters $C_i\subseteq V, |C_i|= p$, and the set of optical switches $SW$ into two subsets, \textit{unit switches} ($SW^{\cup}$) and \textit{cross switches} ($SW^{\times}$), such that:
$SW = SW^{\cup} \cup SW^{\times}$, where $|SW|= \frac{k}{2} \cdot\left\lceil {n}/{p}\right\rceil^2$ if $k$ is the maximum vertex degree of $N_t\in\mathcal{N}$\footnote{W.l.o.g., we assume that OCS switches are full-duplex, i.e., provide bidirectional communication. Note that this assumption can be removed by replacing each full-duplex switch with two simplex switches.}. If $N_t$ is a BST, then
\begin{equation}
|SW|=(\lceil {n}/{p}\rceil^2 + \lceil{n}/{p}\rceil).\label{eq:numBSTswitches}
\end{equation}

\textbf{OpticNet port mapping:} Nodes are assigned to clusters as follows: $\forall (u,v)\in V\times V$, if $u<v$ then $u\in C_i, v\in C_j$, where:
$i = \min\left(\left\lfloor{u}/{p}\right\rfloor, \left\lfloor {v}/{p}\right\rfloor\right)$, $j = \max\left(\left\lfloor {u}/{p}\right\rfloor, \left\lfloor {v}/{p}\right\rfloor\right)$.
Each cluster $C_i$ is assigned to a group of $k$ switches in $SW^{\cup}$, and each distinct pair of clusters $(C_i, C_j), i \neq j$ is assigned to a group of $2k$ switches of type $SW^{\times}$.
The in$\leftrightarrow$out port-pairs are defined according to the type (unit or cross) of each OCS switch $sw_x\in SW$, as $sw^{\cup}_{i\cdot \lceil k / 2 \rceil + x} : C_i\times C_i, ~\forall ~0\leq i < C, 0 \leq x < \lceil k / 2 \rceil$, $sw^{\times}_{idx(i,j)+ x} : C_i \times C_j, ~\forall~ 0\leq i<j < C, 0 \leq x < k$, where the \textit{cluster-index} is defined as $idx(i, j) = \frac{k}{2}\cdot i \cdot (2 \cdot C - 1 - i) +  k \cdot (j - i - 1)$.

For the special case of a BST topology, the definitions above become:
\begin{eqnarray}
sw^{\cup}_{2i + x} &:& C_i\times C_i, ~\forall ~0\leq i < C, 0 \leq x < 2,\nonumber\\
sw^{\times}_{idx(i,j)+ x} &:& C_i \times C_j, ~\forall~ 0\leq i<j < C, 0 \leq x < 2,\nonumber\\
idx(i, j) &=& i \cdot(2 \cdot\left\lceil\frac{n}{p}\right\rceil - i - 1) +  2\cdot (j - i - 1).~~\label{eq:clusterindexingmirrored}
\end{eqnarray}

Figure \ref{fig:ExMirroredBST} illustrates a port mapping example for a BST network with $n=8$, $p=4$, $C=2$, and $|SW|=8$.

\section{Opportunistic Optical Link Scheduling}
\label{sec:ools}

\begin{figure}[t]
\centering
\includegraphics[width=\columnwidth]{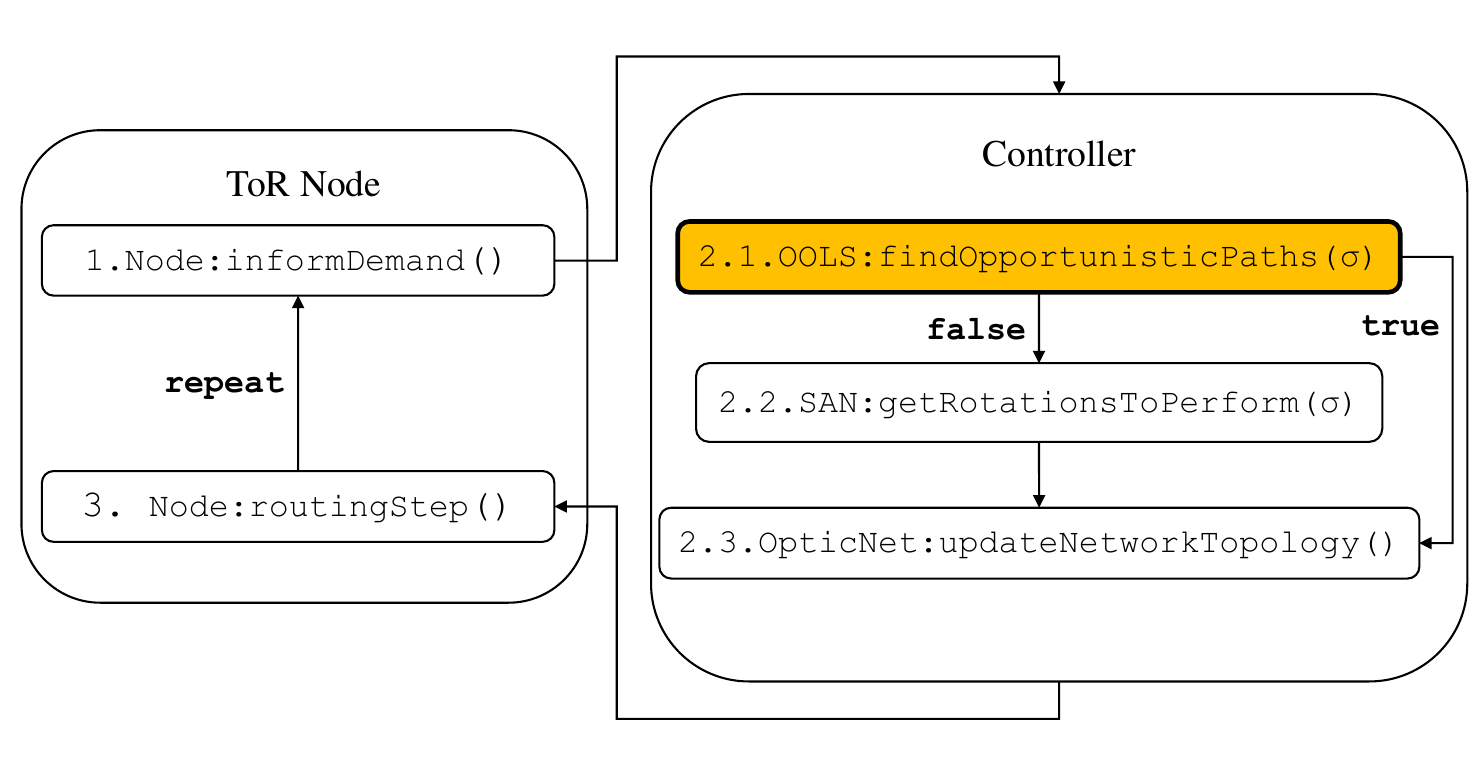}
\caption{OOLS and OpticNet integration.}
\label{fig:diagramaEstados}
\end{figure}

Opportunistic Optical Link Scheduling (OOLS) is a set of routines designed to integrate with a TMT network architecture \cite{spaa21lazy} and leverage idle optical switch ports to create shortcut routing paths opportunistically. In particular, we implemented OOLS within the OpticNet framework \cite{tcc24}.

\textbf{Augmented network topology:} OOLS adds opportunistic links to the SAN topology as long as there are available OCS ports for that. Given a network $N_t\in\mathcal{N}$ of maximum vertex degree $k$, the maximum vertex degree of the augmented network becomes $kC$. For the special case of a BST, this number is $2C+2$, where $C=\lceil n/p\rceil$ is the number of clusters. We refer to the augmented network as $N_{Ut}$. 

\textbf{Extended notation:} Consider a round $t\geq 1$. OOLS creates an additional matching $M_{HL}(x,t)$ ($\nsubseteq E(N_t)$) for each switch $sw_x\in SW$. Note that $N_{Ut} = \cup_{x=1}^{n'} M_{U}(x, t)$, where $M_{U}(x, t) = M(x, t) \cup M_{HL}(x, t)$. We refer to the set of opportunistic link matchings at a time $t$ as $\mathcal{M}_{HLt}$. The set of all matchings at a time $t$ is denoted by $$\mathcal{M}_{Ut} = \mathcal{M}_{t} \cup \mathcal{M}_{HLt}.$$ 

Finally, we denote by $\mathcal{L}(s,d)$ and $\mathcal{L}(s,w,d)$ the opportunistic paths of one hop and two hops, connecting nodes $s$ to $d$, $\sigma(s,d)\in\sigma, w\in V$; and we denote by $(t-1), t'$ and $(t+1)$ the previous, current, and subsequent (simulation) rounds, respectively.  


\textbf{Integration with OpticNet:} Figure \ref{fig:diagramaEstados} illustrates the integrated workflow of OOLS and OpticNet. The simulation proceeds in synchronous rounds, during which the network topology can be adjusted and messages can traverse at most one hop.
Each round is further divided into three time-slots.
Leaf-layer nodes generate the demand (\textbf{Step 1:} \texttt{Node:InformDemand}) and route messages ($\sigma(s,d)\in\sigma$) (\textbf{Step 3:} \texttt{Node:routingStep}). A \textit{centralized controller} mediates the communications between the leaf (ToR) and spine (OCS) switch layers (\textbf{Step 2:} \texttt{OpticNet:controllerStep}). It maintains an up-to-date global view of the topology ($N_t, \forall t\geq 1$), manages the dynamic mapping of links ($(u,v)\in E_t)$) to matchings ($M_U(x,t)\in \mathcal{M}_{Ut}$) and matchings to optical switches ($sw_x\in SW$) (\textbf{Step 2.3:} \texttt{OpticNet:updateNetworkTopology()}), and keeps track of link adjustments ($M_U(x,t) \setminus M_U(x,t-1), \forall sw_x\in SW$) (\textbf{Step 2.2:} \texttt{SAN:getRotationsToPerform($\sigma$)}). OOLS acts in the beginning of the controller step (\textbf{Step 2.1:} \texttt{OOLS:findOpportunisticPaths($\sigma$)}).

\textbf{\texttt{OOLS:findOpportunisticPaths} (Algorithm \ref{alg:updateConn1and2Hops}):} 
In each round $t$, for each node $s\in V$, carrying a message $\sigma(s,d)\in \sigma$, \texttt{OOLS} verifies if there is any path to the destination node $d\in V$, of length one or two, in the augmented network topology $N_{Ut}$. If there is, routing of message $\sigma(s,d)$ is allowed, i.e., the message will be sent in Step $3$ of the current round. If not, then it attempts to create a one-hop opportunistic link (\texttt{OOLS-1H}). If the attempt is successful, then $\mathcal{L}(s,d)$ is created and the message will be sent in Step 3 of the current round. Else, it tries to create (partially, if existing links can be utilized) a two-hop opportunistic path $\mathcal{L}(s,w,d)$ (\texttt{OOLS-2H}). If successful, the message will be sent in Step $3$ of the current and the following rounds. Else, the message is left to be handled by the SAN algorithm (Step $2$). 

Consider a message $\sigma(s,d)\in\sigma$, in which $s$ belongs to cluster $C_s$ and $d$ to cluster $C_d$. Let $i = \min(s,d)$ and $j = \max(s,d)$. We denote the set of switches associated with the cluster pair $(C_s,C_d)$ by $S_s^d=\{sw_z\in SW^{\cup} |~ i\cdot 4 \leq z < (i + 1) \cdot 4 \},$ if $i=j,$ and
$S_s^d=\{sw_z\in SW^{\times} |~ idx(i,j) \leq z < idx(i,j)+ 4\},$ otherwise ($idx(i,j)$ defined as in Eq. (\ref{eq:clusterindexingmirrored})).

\textbf{\texttt{OOLS-1H}:} First, we look for a one-hop path $\mathcal{L}(s,d)$ in $N_{Ut'}$. If such a path does not exist, then we try to create one:
\begin{itemize}
\item \textbf{Case 1 (RouteSD):} A one-hop path $\mathcal{L}(s,d)$ exists in $N_{Ut}$ if $\exists sw_z\in S_s^d ~|~ (s, d) \in M_U(z,t')$. 
\item \textbf{Case 2 (CreateSD):} $\mathcal{L}(s,d)$ is created if $\exists sw_z\in S_s^d ~|~ \forall x\in C_d: (s,x)\notin M_U(z,t') \land 
\forall x\in C_s: (x,d)\notin M_U(z,t').$ 
~~
\end{itemize}    

\begin{figure*}[htbp]
\centering
\includegraphics[width=0.32\textwidth]{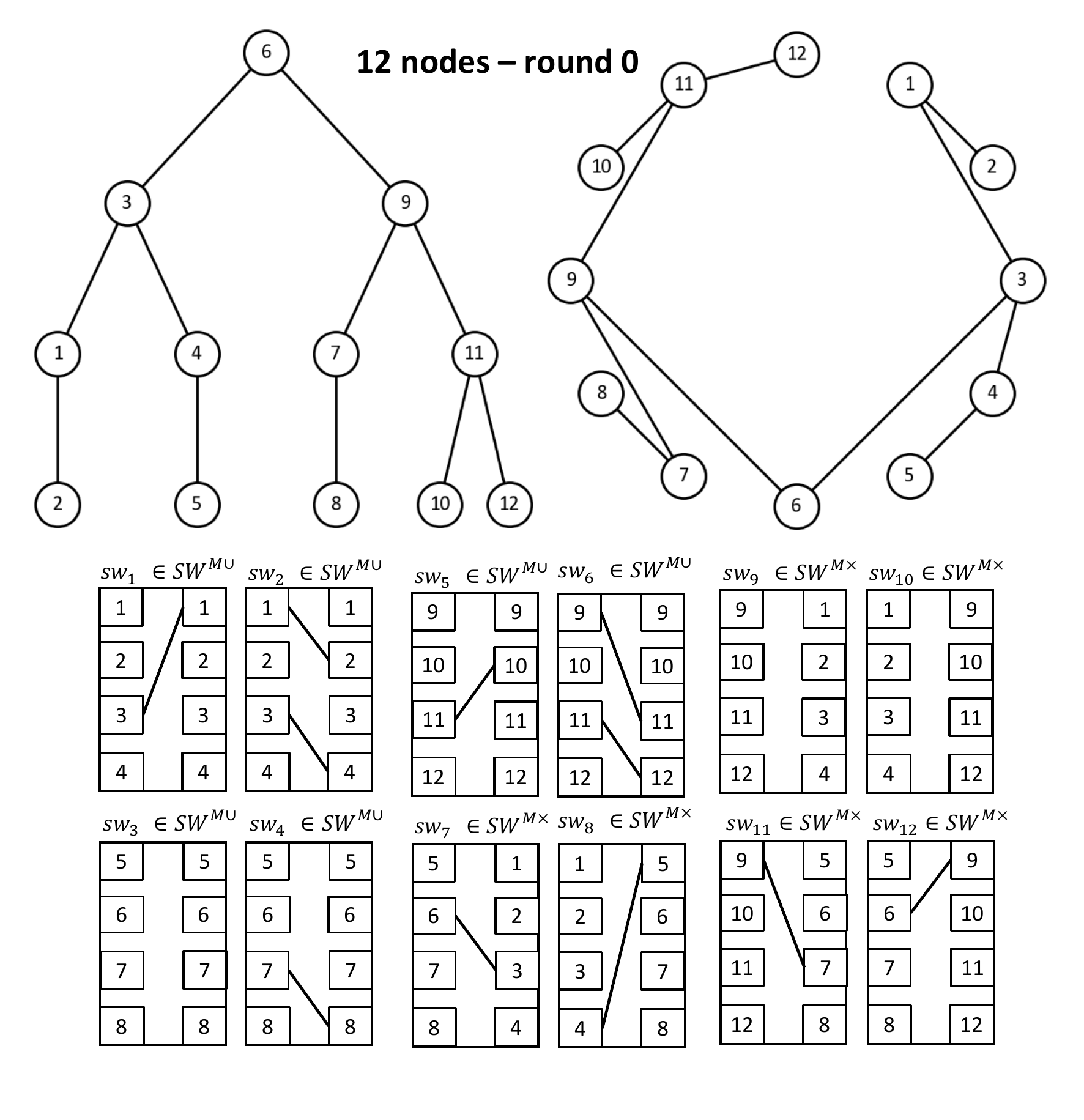}
\hfill
\centering
\includegraphics[width=0.33\textwidth]{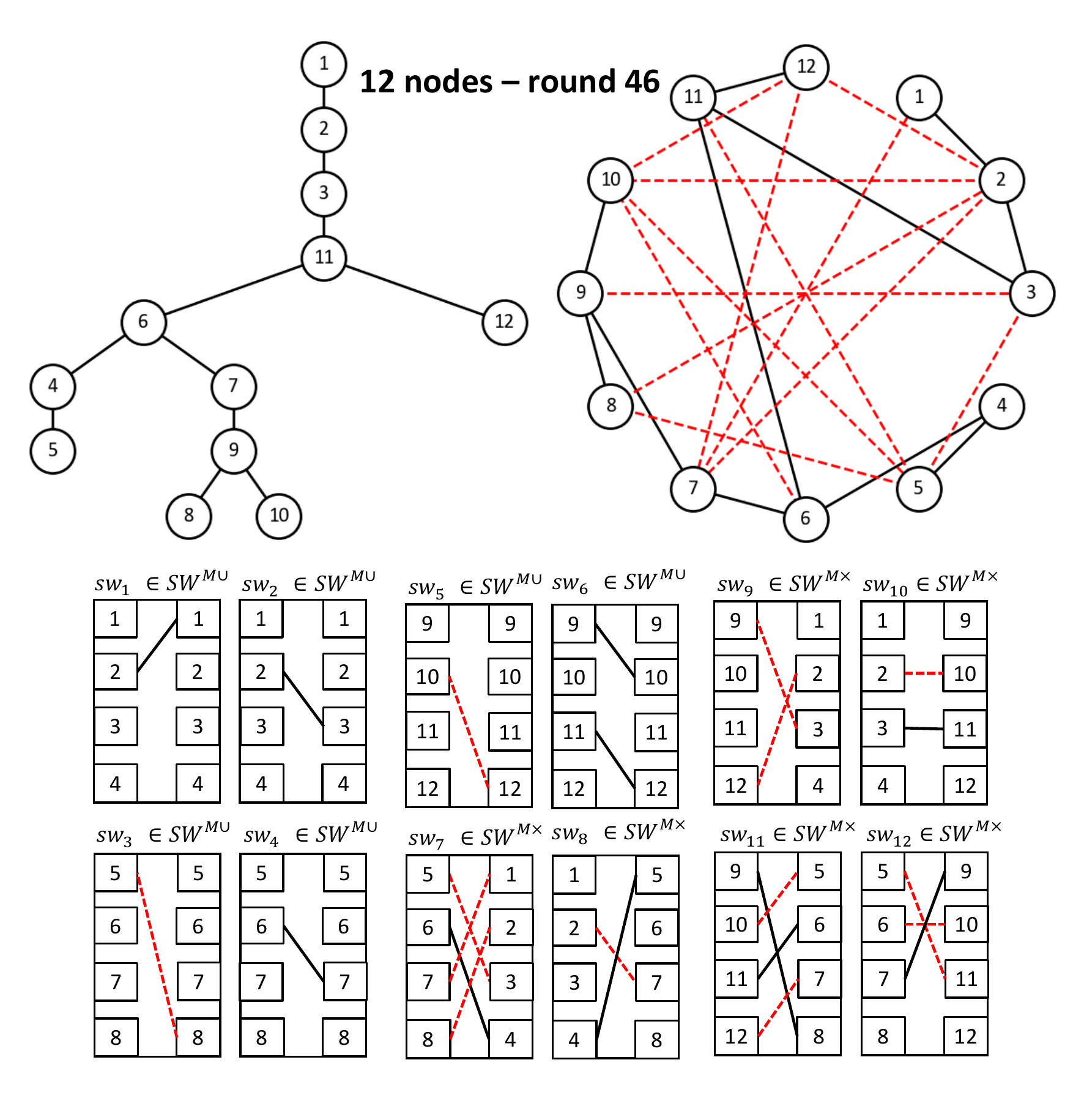}
\centering
\includegraphics[width=0.33\textwidth]{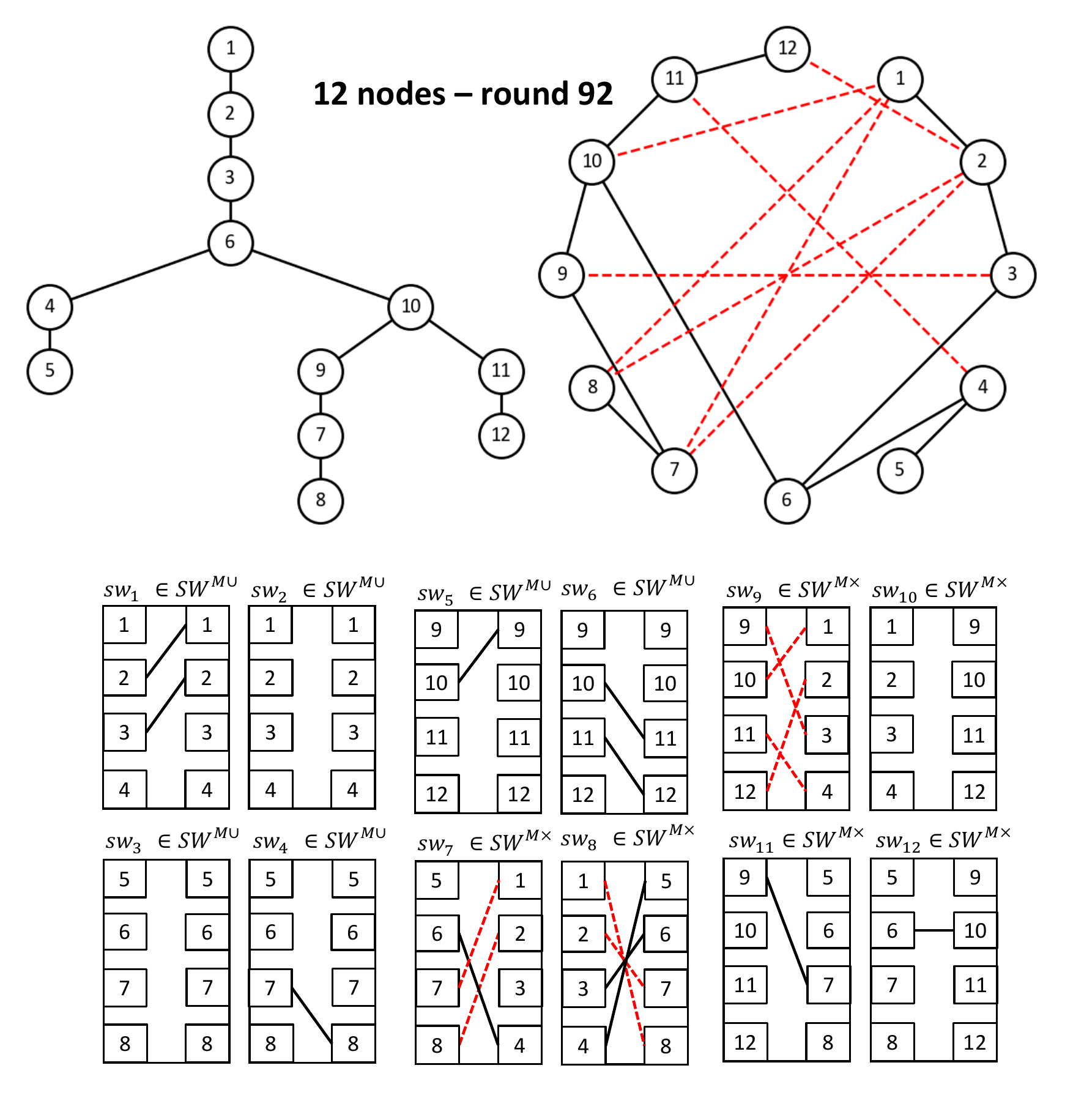}
\caption{\texttt{OOLS-1H} execution: DiSplayNet, $n=12, p=4, C=3,$ $|SW|=12,$ rounds $0$ (first), $46$ (middle), and $92$ (final).}
    \label{fig:12nodetest_1hop}
\end{figure*}

\textbf{Example:} Figure~\ref{fig:12nodetest_1hop} illustrates the execution of \texttt{OOLS-1H} for the DiSplayNet SAN with $n=12, p=4, C=3,$ $|SW|=12,$ in round $46$. Observe that Case 1 (RouteSD) evaluates to \texttt{true} when the message being processed is, e.g., $\sigma(10,12)$. Case 2 (CreateSD) evaluates to \texttt{true} when the message being processed is $\sigma(4,9)$ (or $\sigma(9,4)$). Note that the maximum vertex degree of the augmented network topology is upper bounded by $2C+2=8$. 


\begin{algorithm}[H]
\caption{\texttt{OOLS:findOpportunisticPaths}($\sigma$)}
\label{alg:updateConn1and2Hops}
\begin{algorithmic}[1]
\If {RouteSD} \texttt{\%OOLS-1H, OOLS-2H}
    \State allow $s$ to forward $\sigma(s,d)$ to $d$
    \State \Return true
\EndIf
\If {RouteSWD} \texttt{\%OOLS-2H}
    \State allow $s$ to forward $\sigma(s,d)$ to $w$
    \State \Return true
\EndIf
\If {CreateSD}\texttt{\%OOLS-1H, OOLS-2H}
    \State add $(s,d)$ to $\mathcal{M}_{HLt}$
    \State \Return true
\EndIf
\If {LR1Hop} \texttt{\%OOLS-1H with LR}  
    \State remove links conflicting with $(s,d)$ from $\mathcal{M}_{HLt}$
    \State add $(s,d)$ to $\mathcal{M}_{HLt}$
    \State \Return true
\EndIf
\If {CreateSWOrCreateWD} \texttt{\%OOLS-2H}
    \State add $(s,w)$ or $(w,d)$ to $\mathcal{M}_{HLt}$
    \State \Return true
\EndIf
\If {CreateSWCreateWD} \texttt{\%OOLS-2H}
    \State add $(s,w)$ and $(w,d)$ to $\mathcal{M}_{HLt}$
    \State \Return true
\EndIf
\If {LR2Hops} \texttt{\%OOLS-2H with LR}
    \State remove links conflicting with $(s,w)$ or $(w,d)$ from $\mathcal{M}_{HLt}$
    \State add $(s,w)$ and $(w,d)$ to $\mathcal{M}_{HLt}$
    \State \Return true
\EndIf
\State \Return false
\end{algorithmic}
\end{algorithm}

\textbf{\texttt{OOLS-2H}:} If \texttt{OOLS-1H} returns \texttt{false} for a given message $\sigma(s,d)\in\sigma$, then we look for a two-hop path $\mathcal{L}(s,w,d)$ in $N_{Ut'}$. If such a path does not exist, then we try to reuse an already existing link $(s,w)$ or $(w,d)$ from $N_{Ut'}$ and create the remaining half of the path. If no such opportunity is found, we search for a relay node $w\in V$ and try to create both hops from scratch:
\begin{itemize}
    \item \textbf{Case 3 (RouteSWD):} A two-hop path $\mathcal{L}(s,w,d)$ exists in $N_{Ut'}$ if
    $\exists w \in V~|~  
    (w \neq s \land w \neq d )\land \newline
    \exists sw_x \in S_s^w ~|~ (s, w) \in M_U(x,t') \land \newline
    \exists sw_y\in S_w^d ~|~ (w, d) \in M_U(y, t')$.
    \item \textbf{Case 4 (CreateSWOrCreateWD):} There is a cluster $C_w$ and a relay node $w \in C_w$, such that one hop of the $\mathcal{L}(s,w,d)$ path already exists in $N_{Ut'}$ and the other hop can be created. The insertion of the opportunistic link is successful if:
    $\exists w \in V~|\newline
    (w \neq s \land w \neq d )\land \newline
    (\exists sw_x\in S_s^w ~|~ (s, w) \in M_U(x,t')\land \newline
    \exists sw_z\in S_w^d ~|~  
    (\forall v\in C_d: (w,v)\notin M_U(z,t')) \land \newline 
    (\forall v\in C_w: (v,d)\notin M_U(z,t')) \lor \newline\newline  
    \exists sw_y\in S_w^d ~|~ (w, d) \in M_U(y,t') \land \newline 
    \exists sw_z\in S_s^w ~|~
    (\forall v\in C_w: (s,v)\notin M_U(z,t')) \land \newline 
    (\forall v\in C_s): (v,w)\notin M_U(z,t'))$.
    \item \textbf{Case 5 (CreateSWCreateWD):} There is a cluster $C_w$ and a relay node $w \in C_w$, such that the $\mathcal{L}(s,w,d)$ path can be created. The insertion of the new link is successful if:
    $\exists w \in C_w, C_w\in topC(C_s) ~|~\newline
    (w \neq s \land w \neq d )\land \newline 
    \exists sw_x\in S_s^w ~|~ 
    (\forall v\in C_w: (s,v)\notin M_U(x,t')) \land \newline 
    (\forall v\in C_s: (v,w)\notin M_U(x,t')) \land \newline
    \exists sw_y\in S_w^d ~|~ 
    (\forall v\in C_d: (w,v)\notin M_U(y,t')) \land \newline 
    (\forall v\in C_w: (v,d)\notin M_U(y,t'))$.
\end{itemize}
Note that, when combining the Cases $1$ through $5$, Case $3$ (RouteSWD) should be executed before Case $2$ (CreateSD), to avoid creating unnecessary one-hop opportunistic links. 

\begin{figure}[t]
\centering
\begin{subfigure}[b]{0.67\columnwidth}
\includegraphics[width=\columnwidth]{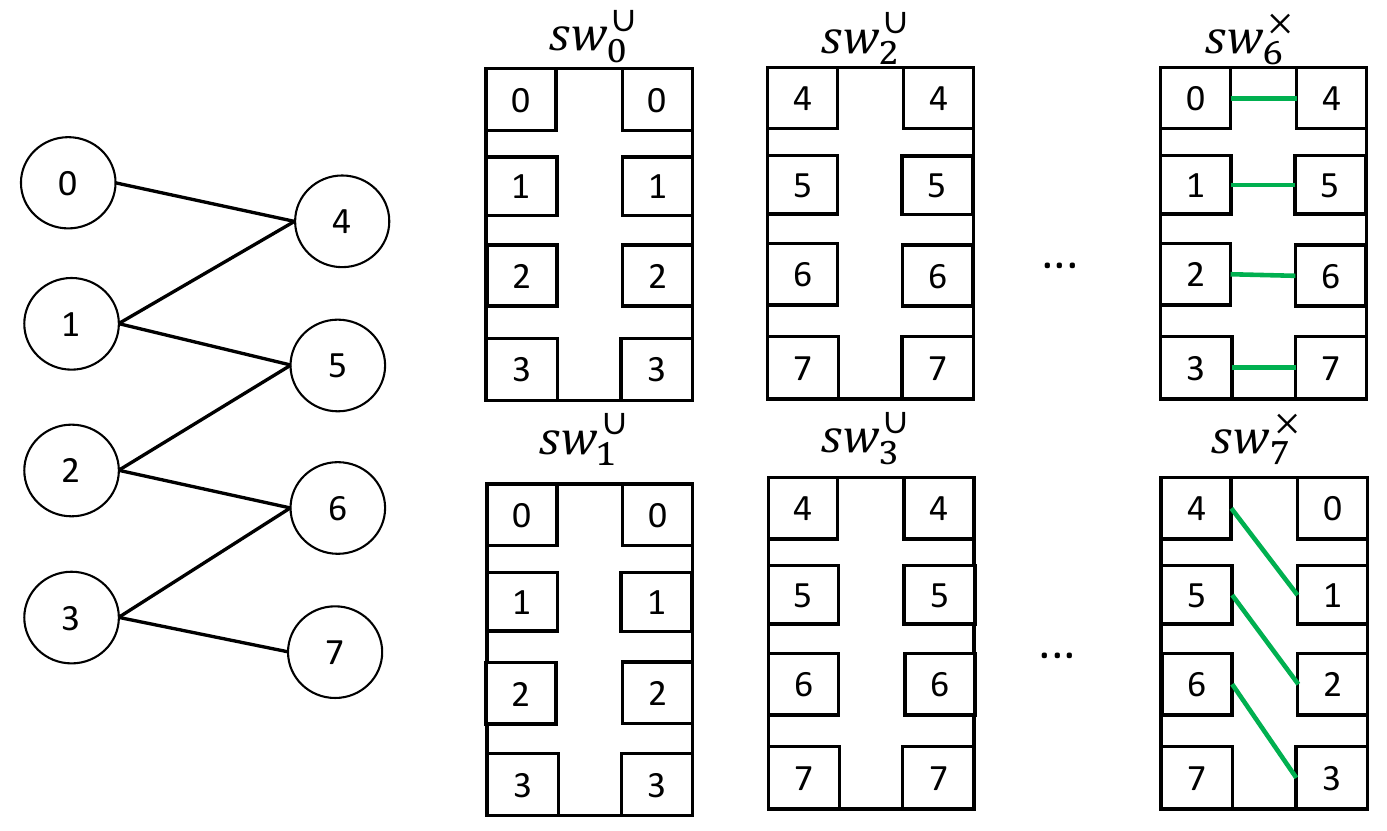}
\caption{Only SAN links (BST w/ $n=8, p=4, C=2, |SW|=8$)}
\label{fig:ExMirroredBST}
\end{subfigure}
\hfill
\begin{subfigure}[b]{0.3\columnwidth}
\includegraphics[width=\columnwidth]{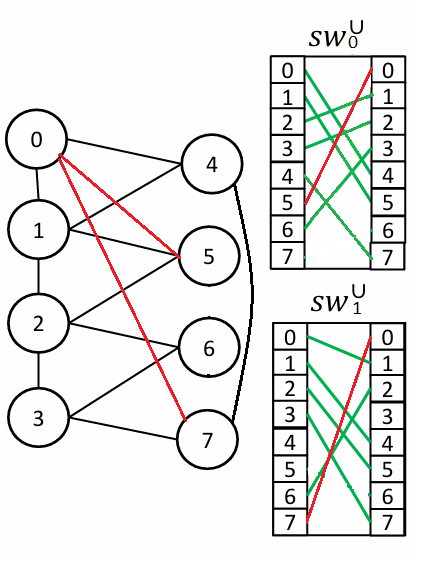}
\caption{\texttt{OOLS-2H} ($n=8, p=8, |SW|=2$)}
\label{fig:add2HopDuplex}
\end{subfigure}
\caption{OpticNet link-to-port mapping: (a) Only SAN links; (b) Augmented topology w/ 2-hop OOLS path $\mathcal{L}(5,0,7)$.}
\end{figure}

\textbf{Example:} Figure~\ref{fig:add2HopDuplex} illustrates the execution of \texttt{OOLS-2H} for the same network shown in Figure~\ref{fig:12nodetest_1hop}, right before adding the opportunistic path $\mathcal{L}(5,0,7)$. Observe that Case $1$ (RouteSWD) evaluates to \texttt{true} for the message $\sigma(1,6)$ ($w=2$). Case $4$ (CreateSWOrCreateWD) evaluates to \texttt{true} for $\sigma(7,1)$ ($w=0$). Case $5$ (CreateSWCreateWD) evaluates to \texttt{true} for $\sigma(5,7)$ ($w=0$). 


OOLS never overwrites non-opportunistic links ($(u,v)\in N_t$), and whenever a SAN link adjustment (rotation) occurs, new (BST) links always overwrite any opportunistic links that occupy the same OCS ports. Moreover, old opportunistic links are overwritten by newer opportunistic links using cache replacement policies, such as Least Recently Used (LRU) or Least Frequently Used (LFU).

\textbf{\texttt{OOLS with LR}:}
When OOLS fails to find free OCS ports to add an opportunistic link (within the limited search space of $topC$ candidate switches), it removes older conflicting opportunistic links using a cache replacement policy. Link Replacement (LR) is activated in two cases, described below:
\begin{itemize}
    \item \textbf{Case 6 (LR1Hop):} For each $sw_z\in S_s^d,$ we consider the set of all opportunistic links $(s,x) \in M_{HL}(z,t')$ incident to $s$ (when the $d$ output port is free) or $d$ ($(x,d)\in M_{HL}(z,t')$, when the $s$ input port is free). We remove the one with the lowest LR score (LFU or LRU). The insertion of the new link $\mathcal{L}(s,d)$ is successful iff: 
    $\exists sw_z\in S_s^d | \newline
    (\forall x\in C_d: (s,x)\notin M_U(z,t')\setminus \{l_{sx}\} \land \newline
    \forall x\in C_s: (x,d)\notin M_U(z,t')) \lor \newline
    (\forall x\in C_d: (s,x)\notin M_U(z,t') \land \newline
    \forall x\in C_s: (x,d)\notin M_U(z,t')\setminus \{l_{xd}\}),$ where:\\
    $l_{xd} = \arg\min_{M_{HL}(z,t'), sw_z\in S_s^d}{(LR(x,d))}, \newline
    l_{sx} = \arg\min_{M_{HL}(z,t'), sw_z\in S_s^d}{(LR(s,x))}.$
    \item \textbf{Case 7 (LR2Hops):} 
    Given the set of candidate clusters $C_w\in topC(C_s)$, $\forall w\in C_w,$ we search for a switch $sw_y\in S_s^w$, such that either both (input and output) ports $(s,w)$ are free (i.e. $\notin M_U(z,t;)$), or become free by removing one opportunistic link (either $l_{s,x}$ or $l_{xw}$) from $M_{HL}(z,t')$; then we repeat the search in $S_w^d$ for the second hop $(w,d)$. Consider the set of all switch pairs $(sw_y, sw_z)$ that satisfy the condition above. Then, the new link will be inserted into the pair with the lowest combined score ($LR(l_s)+LR(l_d)$), where
    $l_s$ ($l_d$) is either $l_{sx}$ ($l_{wx}$) or $l_{xw}$ ($l_{xd}$), depending on which (if any) link was removed from $sw_y$ ($sw_z$), $LR(\emptyset)=0,$ and:\\
    $l_{xd} = \arg\min_{M_{HL}(z,t'), sw_z\in S_w^d}{(LR(x,d))}, \newline
    l_{wx} = \arg\min_{M_{HL}(z,t'), sw_z\in S_w^d}{(LR(w,x))}, \newline
    l_{sx} = \arg\min_{M_{HL}(z,t'), sw_z\in S_s^w}{(LR(s,x))}, \newline
    l_{xw} = \arg\min_{M_{HL}(z,t'), sw_z\in S_s^w}{(LR(x,w))}.$


    The insertion of the new link is successful iff: $\exists w \in C_w, C_w\in topC(C_s) |
    (w \neq s \land w \neq d )\land \newline 
    (
    	\exists sw_y \in S_s^w: (s,w)\in M_U(y,t') \lor\newline 
    	(
    		\forall x\in C_s: (x,w)\notin M_U(y,t')\setminus\{l_{xw}\} \land \newline
    		\forall x\in C_w: (s,x)\notin M_U(y,t') \lor \newline
    		\forall x\in C_s: (x,w)\notin M_U(y,t') \land \newline
    		\forall x\in C_w: (s,x)\notin M_U(y,t')\setminus \{l_{sx}\} \lor \newline
            \forall x\in C_s: (x,w)\notin M_U(y,t') \land \newline
    		\forall x\in C_w: (s,x)\notin M_U(y,t')
    	)
    ) \land \newline\newline
    (
    	\exists sw_z \in S_w^d: (w,d)\in M_U(z,t') \lor \newline 
    	(
    		\forall x\in C_w: (x,d)\notin M_U(z,t')\setminus\{l_{xd}\} \land \newline
    		\forall x\in C_d: (w,x)\notin M_U(z,t') \lor \newline
    		\forall x\in C_w: (x,d)\notin M_U(z,t') \land \newline
    		\forall x\in C_d: (w,x)\notin M_U(z,t')\setminus \{l_{wx}\} \lor \newline
            \forall x\in C_w: (x,d)\notin M_U(z,t') \land \newline
    		\forall x\in C_d: (w,x)\notin M_U(z,t') )).$ 

    ~~
\end{itemize}


\textbf{Example:} Consider an example execution of \texttt{OOLS with LR}, illustrated in Figure~\ref{fig:12nodetest_1hop}, in round $46$. Case LR1Hop evaluates to \texttt{true} for, e.g., the message $\sigma(2,6)$. Case LR2Hop, on the other hand, evaluates to \texttt{true} for $\sigma(10,7)$ (using node $w=3$ as relay).

\textbf{Complexity} Given a message $\sigma(s,d)\in\sigma$, the computational complexity of \texttt{OOLS-1H} (Cases 1,2, and 6) is bounded by $|S_s^d|$, which is $\leq 4,\forall (u,v)\in V\times V$ for the special case of a BST, and $O(k)$ for the more general case of a bounded-degree network family $\mathcal{N}$, where $k$ is the maximum vertex degree in $N_t\in\mathcal{N}$.
The cost of \texttt{OOLS-2H} (Cases 3,4,5 and 7) is dominated by the size of the search space for the relay node $w$, which can belong to one of as many as $C=O(n/p)$ clusters. 
To reduce complexity, for each cluster $C_i, 1\leq i\leq C$, we maintain a priority list $topC(C_i)$ of $<C$ clusters $C_j, j\neq i$, sorted by (non-decreasing) number of occupied links, i.e., $|\{(u,v)\in M_{HL}(z,t'), sw_z\in S_u^v, u\in C_i, v\in C_j\}|$. 
For a message $\sigma(s,d)$, we limit the search space to at most $\min(C,topC)$ candidate clusters $C_w\in topC(C_s), s\in C_s$, where $C_s$ is the cluster of $s$. From each of these (least occupied) clusters $C_w$, we select (uniformly at random) a subset of at most $\min(p,topC)$ nodes $w\in C_w$, and check the feasibility conditions (Cases 3, 4, 5, 7) for all switches in $S_s^w$ and $S_w^d$, which runs in time $O(|S_s^w|+|S_w^d|)=O(1)$ for BST, and $O(k)$ for a bounded-degree topology). If the insertion of the two-hop opportunistic link $\mathcal{L}(s,w,d)$ is successful, we update the priority list $topC(C_s)$.
Putting everything together, we have
$\min(C,topC)\cdot\min(p,topC)\cdot O(k) + O(\log(C))$. For the special case of BST, with $|S_u^v|\leq 4, \forall (u,v)$ and $topC=O(1)$, the cost of \texttt{OOLS-2H} is upper bounded by $O(k + \log(C)) = O(\log(\frac{n}{p}))$ per message. 

\textbf{Concurrency:} To handle concurrent requests for link adjustments, OpticNet adopts an optimistic approach, i.e., conflicts are dealt with upon occurrence. In order to ensure deadlock and starvation freedom, concurrent requests are executed according to a \textit{priority}. Given two messages~$\sigma_i\in\sigma$ and~$\sigma_j\in\sigma$, we say that~$\sigma_i$ has a higher priority than~$\sigma_j$ if~$b_i < b_j$. 
The controller node handles prioritization rules among nodes participating in each (routing and link adjustment) operation to ensure they don't join in conflicting operations. To prevent \textit{multi-round adjustments} (e.g. a \textit{zig-zag} splay) from being interrupted by messages with higher priority, OpticNet handles conflicts around such adjustments by assigning maximum priority to messages related to them, and nodes are never locked for more than one round. 
OOLS adopts and extends this strategy, in line with some related work \cite{2021TCCDisplaynet, 2022COMNETCBNet}. To prevent routing inconsistency, nodes participating in \texttt{OOLS-1H} operations are locked for the current round, and nodes involved in \texttt{OOLS-2H} are locked for the current and the subsequent rounds.



\section{Evaluation}
\label{sec:evaluation}

\textbf{Baselines:} To evaluate the performance of OOLS, we used the OpticNet\footnote{\url{https://github.com/caiocaldeira3/OpticalNet}} simulation framework and the following SAN algorithms as baselines:
\begin{itemize}
    \item DSN: SemiDiSplayNet\footnote{\url{https://github.com/caiocaldeira3/SemiDisplayOpticNet}}, an enhanced version of DisplayNet \cite{displaynet} with additional splay rotations. DisplayNet is a distributed and concurrent version of SplayNet \cite{splaynet};
    \item CBN: CBNet \cite{2022COMNETCBNet}, a distributed and concurrent generalization of CBTree \cite{cbtree}.
\end{itemize}

DiSplayNet is the most dynamic SAN in terms of network reconfiguration. Given a message $\sigma(s,d)$, the source node $s$ performs splaying rotations until $d_{e}(s,d)=1$ and only then forwards the message content to the destination node $d$. In SemiDisplayNet, there are additional top-down semi-splay rotations, some of which also involve routing. CBNet, on the other hand, uses locally stored counters, $s$ decides whether to perform splaying or to forward the message to an intermediate node $v$ along the path to $d$. In the latter case, $v$ becomes the carrier of $\sigma(s,d)$ and repeats the procedure until the message reaches the destination $d$. On average, CBNet performs a much lower number of splaying operations, while preserving a similar total communication cost, compared to SemiDiSplayNet.

\textbf{OOLS versions:} We implemented the following variations of OOLS\footnote{\url{https://github.com/marcostiagofh/OOLS-CBN}},\footnote{\url{https://github.com/marcostiagofh/OOLS-DSN}}:
\begin{itemize}
    \item OOLS-1H(DSN): the DSN-based \textit{one-hop} OOLS; 
    \item OOLS-1H(CBN): the CBN-based \textit{one-hop} OOLS;
    \item OOLS-2H(DSN): the DSN-based \textit{two-hop} OOLS;
    \item OOLS-2H(CBN): the CBN-based \textit{two-hop} OOLS;
    \item OOLS-2H(DSN,LRU):  the DSN-based \textit{two-hop} OOLS with LRU cache (link) replacement;
    \item OOLS-2H(DSN,LFU):  the DSN-based \textit{two-hop} OOLS with LFU cache replacement;
\end{itemize}

\textbf{Metrics:} We used the following metrics to evaluate the performance of the simulated algorithms:
\begin{itemize}
    \item \textit{Throughput} (defined as the total
number of messages delivered per round);
    \item \textit{Makespan} (defined in Eq. (\ref{eq:makespan}));
    \item \textit{Total work} (defined in Eq. (\ref{eq:total_cost}));
    \item \textit{Switch activation}: the percentage of simulation rounds, relative to the makespan, in which an OCS switch performed either a routing or a reconfiguration operation;
\end{itemize}



\begin{table}[t]
\centering
\caption{Workload parameters.}
\resizebox{0.95\columnwidth}{!}{%
\begin{tabular}{|l|c|c|c|c|}
\hline
\textbf{Workload} & $T(\sigma)$ & $NT(\sigma)$ & $n$ & $m$ \\ \hline
pFabric (Trace 0.5)~\cite{pfabric} & $0.431$ & $0.986$ & $144$ & $1,000,000$ \\ \hline
Facebook~\cite{Roy:2015} & $1.000$ & $0.904$ & $367$ & $1,000,000$  \\ \hline
HPC (Multigrid) & $0.535$ & $0.820$ & $1024$ & $1,000,000$  \\ \hline
Bursty~\cite{trace-complexity} & $0.400$ & $1.000$ & $128$ & $10,000$ \\ \hline
Skewed~\cite{trace-complexity} & $1.000$ & $0.400$ & $128$ & $10,000$  \\ \hline
ProjecToR~\cite{projector} & $0.997$ & $0.467$ & $128$ & $10,000$  \\ \hline
\end{tabular}
}
\label{tab:workload_parameters}
\end{table}

\textbf{Workloads:} To quantify the locality of reference in a workload, we use the concept of \textit{trace complexity} \cite{trace-complexity}. In this context, we distinguish between temporal ($T(\sigma)$) and spatial components ($NT(\sigma)$) of locality. We used a set of diverse traffic workloads ($\sigma$) with variable length ($m$), network size ($n$), and spatial ($NT(\sigma)$) and temporal ($T(\sigma)$) localities, as listed in
Table~\ref{tab:workload_parameters}.
Note that $NT(\sigma)$ and $T(\sigma)\in [0,1]$, such that values closer to $1$ indicate \textit{lower locality}, and values closer to $0$ indicate \textit{higher locality}. In particular, we have that:
\begin{itemize}
    \item \textbf{ProjecToR}\footnote{\url{https://www.microsoft.com/en-us/research/project/projector-agile-reconfigurable-data-center-interconnect/}} and \textbf{Skewed} have \textit{high spatial and low temporal locality};
    \item \textbf{PFabric} and \textbf{Bursty} have \textit{high temporal and low spatial locality};
    \item \textbf{HPC}\footnote{\url{https://portal.nersc.gov/project/CAL/overview.htm}} has relatively \textit{high temporal and spatial locality}; 
    \item \textbf{Facebook} has \textit{low temporal and spatial locality}.
\end{itemize}

\textbf{Simulation parameters:} The sequence of messages in each workload was simulated following the Poisson distribution with $\lambda=1$ expected messages per time slot. Since each simulation round of OpticNet is comprised of three time slots, plus one time slot for logging, a sequence of $m$ messages takes, on average, $\approx {m}/{1.48}=0.68m$ simulation rounds to be generated (which is also the expected lower bound for the makespan).




\section{Simulation Results}
\label{sec:results}

In this section we present and discuss our simulation results. In Sec. \ref{subsec:results_throughput}, we analyze the throughput and makespan; in Sec. \ref{subsec:results_work}, we analyze the total work (communication cost); in Sec. \ref{subsec:results_cache}, we analyze the cache replacement strategies; in Sec. \ref{subsec:results_cdf_constant_p} and Sec.\ref{subsec:results_cdf_var_p},  we analyze the activity distribution with constant and variable switch size, resp.; and, finally, in Sec. \ref{subsec:results_conclusion} we present some concluding remarks.

\begin{figure*}[h]
\centering
\begin{subfigure}[b]{0.3\textwidth}
\includegraphics[width=\textwidth]
{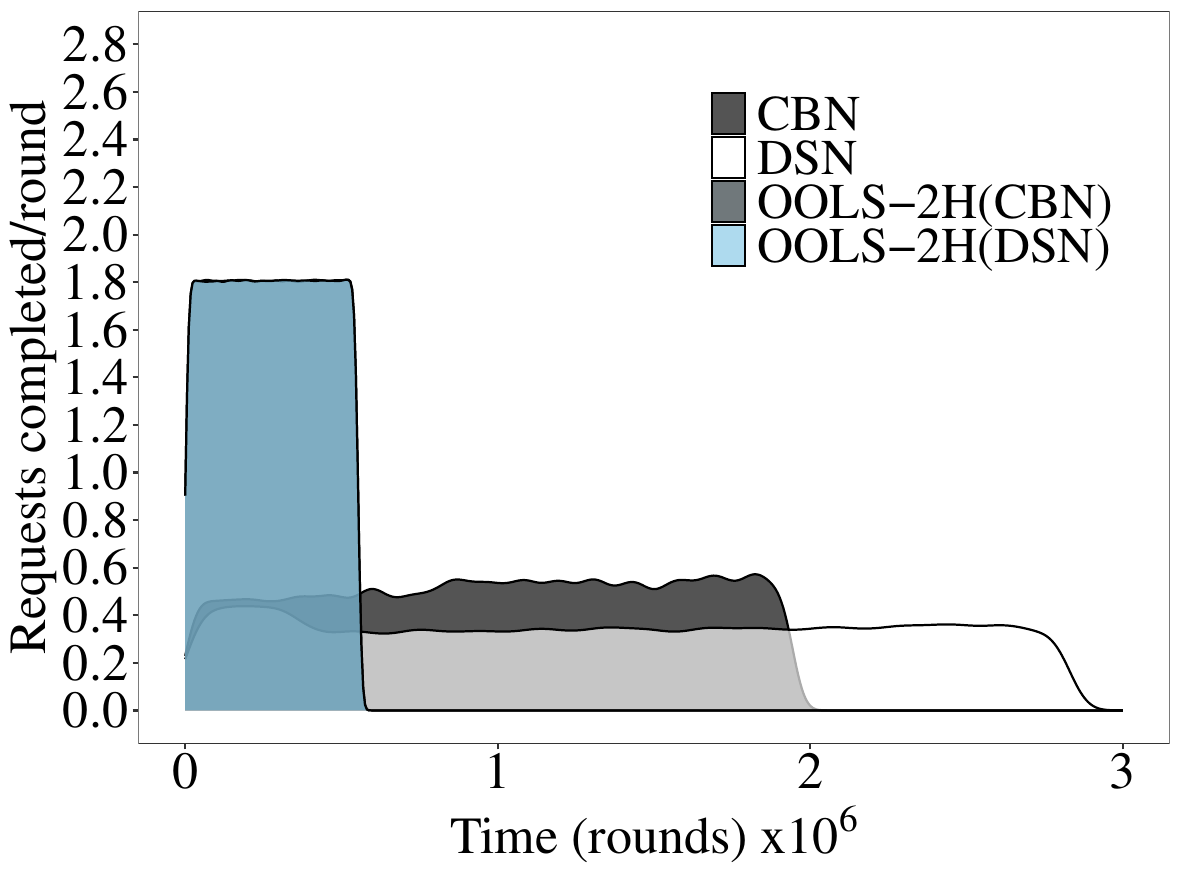}
\caption{pFabric ($n=144, p=9, m=10^6$)}
\label{fig:throughput-pfab}
\end{subfigure}
\hfill
\begin{subfigure}[b]{0.3\textwidth}
\includegraphics[width=\textwidth]{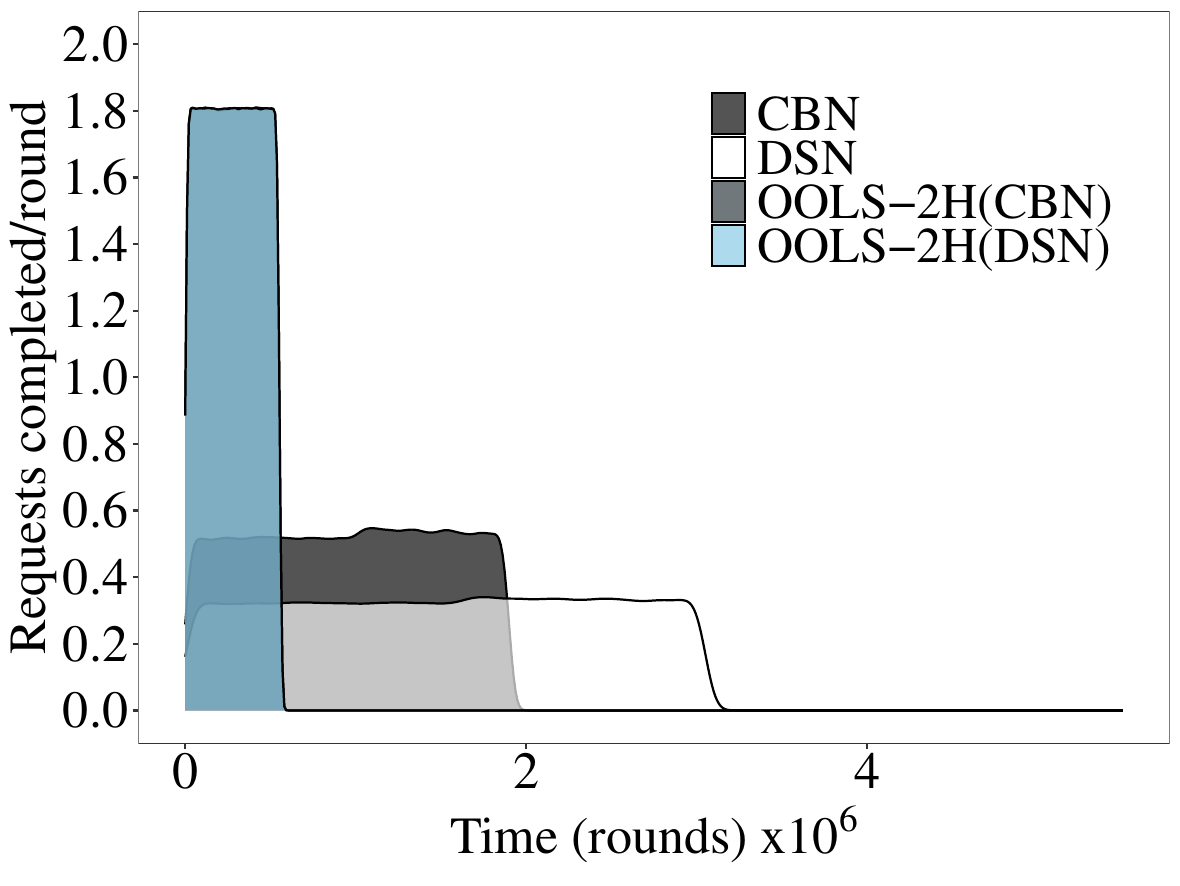}
\caption{Facebook ($n=367, p=23, m=10^6$)}
\label{fig:throughput-facebook}
\end{subfigure}
\hfill
\begin{subfigure}[b]{0.3\textwidth}
\includegraphics[width=\textwidth]
{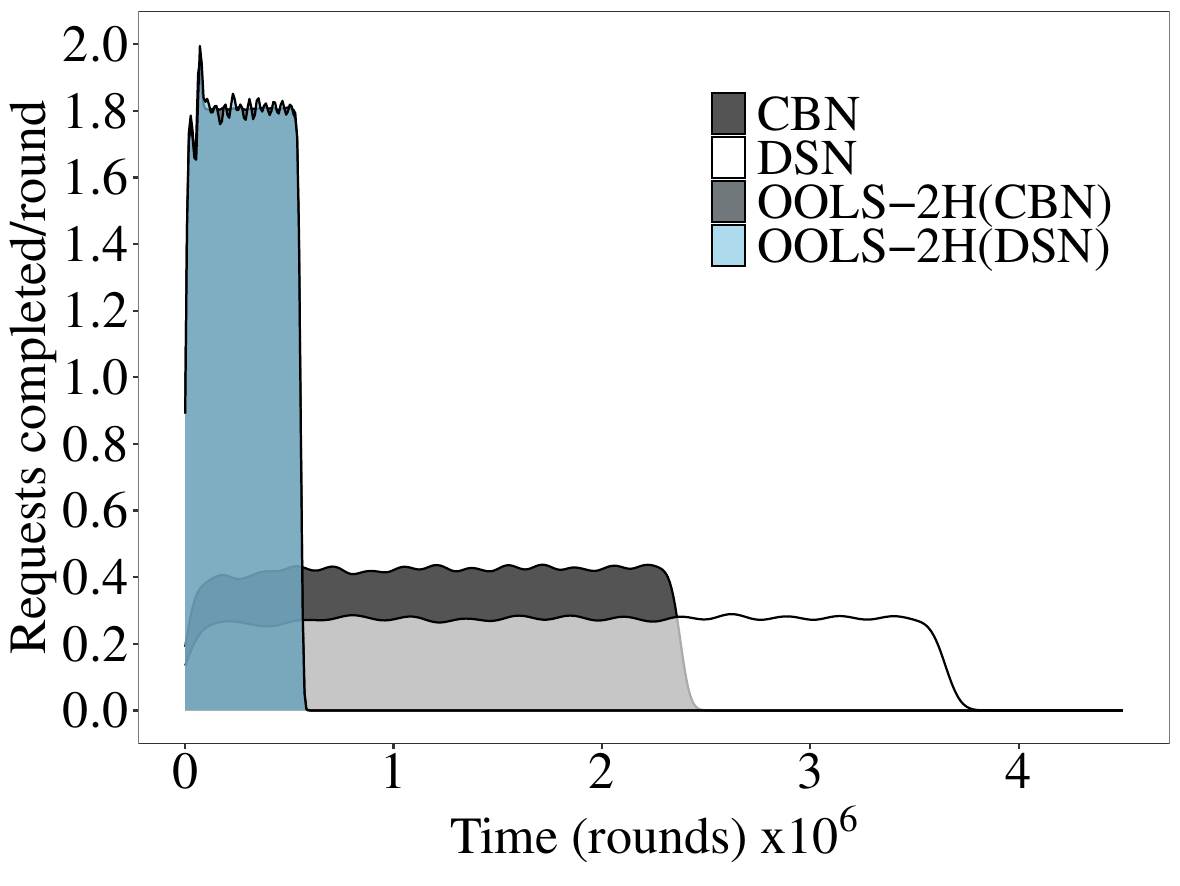}
\caption{HPC ($n=1024, p=64, m=10^6$)}
\label{fig:throughput-hpc}
\end{subfigure}
\hfill
\begin{subfigure}[b]{0.3\textwidth}
\includegraphics[width=\textwidth] {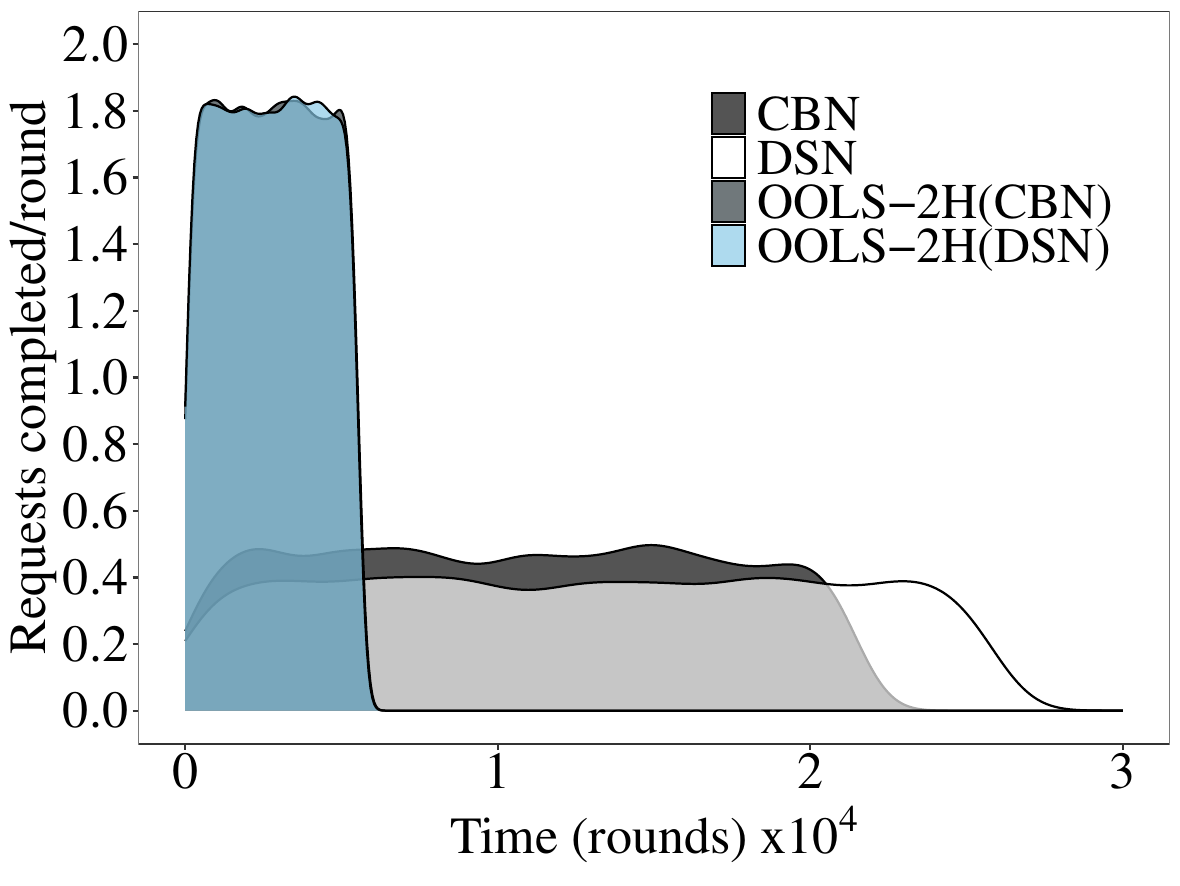}
\caption{Bursty ($n=128,p=8, m=10^4$)}
\label{fig:throughput-bursty}
\end{subfigure}
\hfill
\begin{subfigure}[b]{0.3\textwidth}
\includegraphics[width=\textwidth]{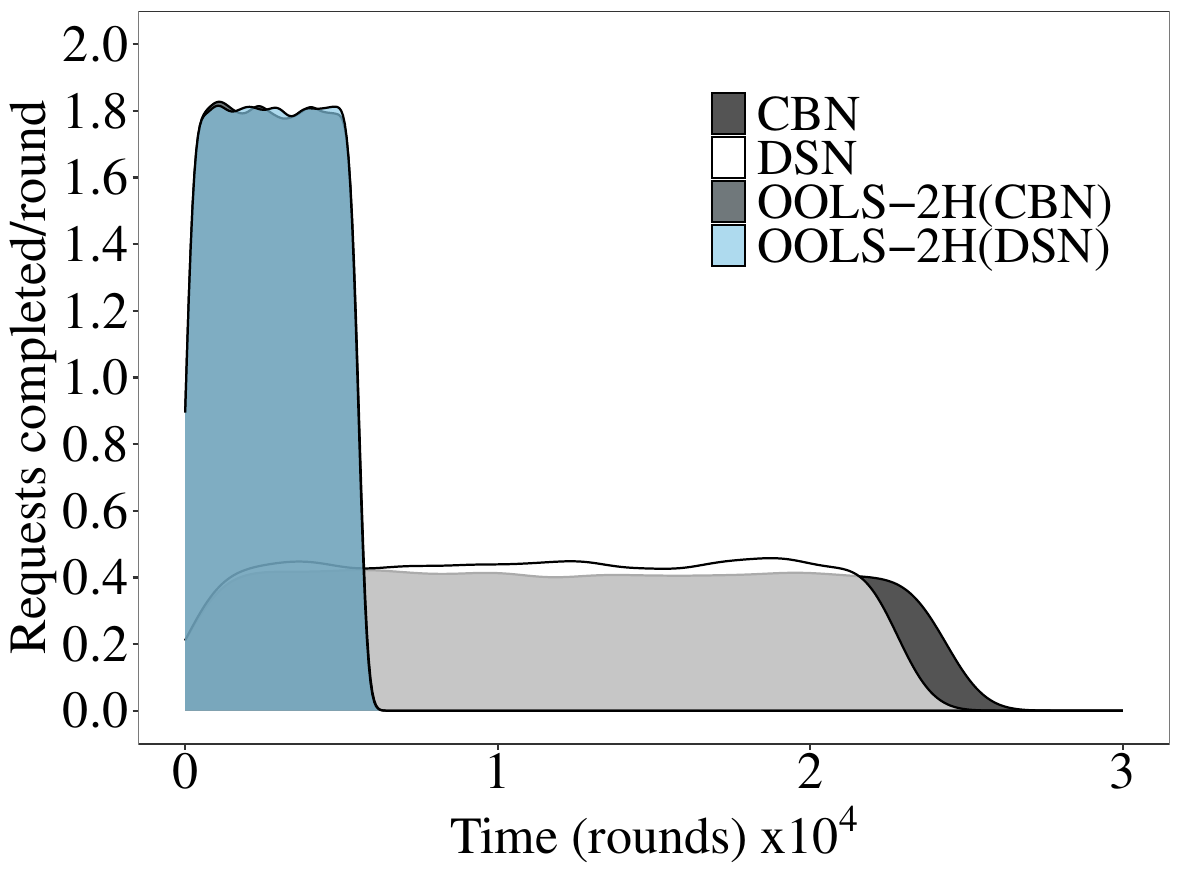}
\caption{Skewed ($n=128,p=8, m=10^4$)}
\label{fig:throughput-skewed}
\end{subfigure}
\hfill
\begin{subfigure}[b]{0.3\textwidth}
\includegraphics[width=\textwidth]
{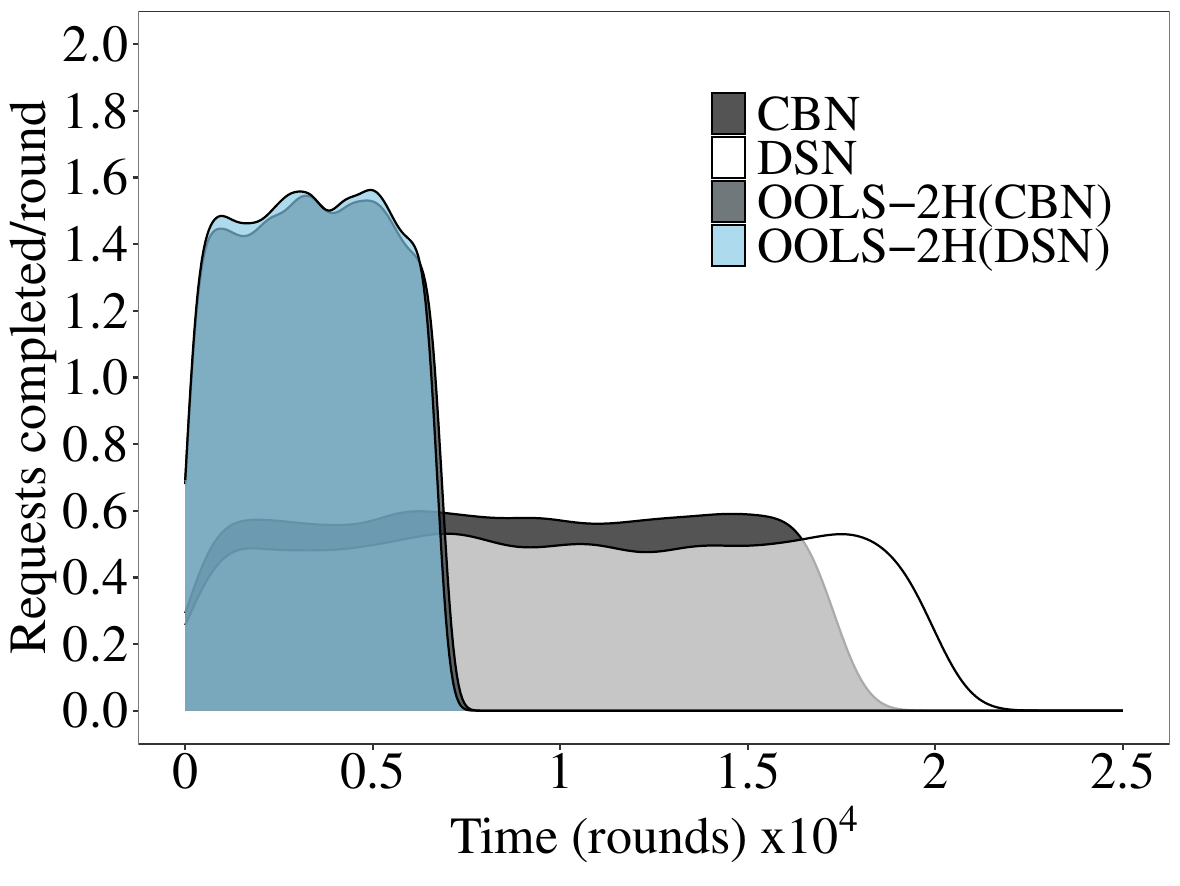}
\caption{ProjecToR ($n=128,p=8, m=10^4$)}
\label{fig:throughput-tor}
\end{subfigure}
\caption{Throughput over all datasets ($ \lceil{n}/{p}\rceil = 16$).}
\label{fig:mu1_throughput_all}
\end{figure*}

\subsection{Throughput and Makespan}
\label{subsec:results_throughput}

In Figure \ref{fig:mu1_throughput_all}, we compare the \textit{throughput}, defined as the total number of messages delivered per round, of the baseline SAN algorithms (CBN and DSN) to the \textit{two-hop} OOLS implemented over those algorithms (OOLS-2H(CBN) and OOLS-2H(DSN), resp.). Note that the \textit{makespan} is represented by the maximum round index on the x-axis. Figures \ref{fig:throughput-pfab} through \ref{fig:throughput-tor} show the results for all datasets. To facilitate the comparative analysis among different datasets, the size of the OCS switch, measured in the number of ports, $p$, has been chosen such that  $\left\lceil \frac{p}{n}\right\rceil \approx 16$, where $n$ is the number of network nodes (ToRs), in all workloads.

We can observe that, in all scenarios, the \textit{makespan} achieved by the baselines is at least $1.7 m$ (CBN) and $2.0 m$ (DSN), where $m$ is the total number of messages exchanged during each simulation. \textit{Two-hop} OOLS, on the other hand, delivers all messages in less than $0.7 m$ rounds in all workloads.
This reveals that OOLS significantly increases the concurrency of the system, relative to the baseline SAN algorithms, which struggle to sustain concurrency due to costly locking operations required during conflicting rotations.

\begin{figure*}[h]
\centering	
\begin{subfigure}[b]{0.32\textwidth}
    \includegraphics[width=\textwidth]{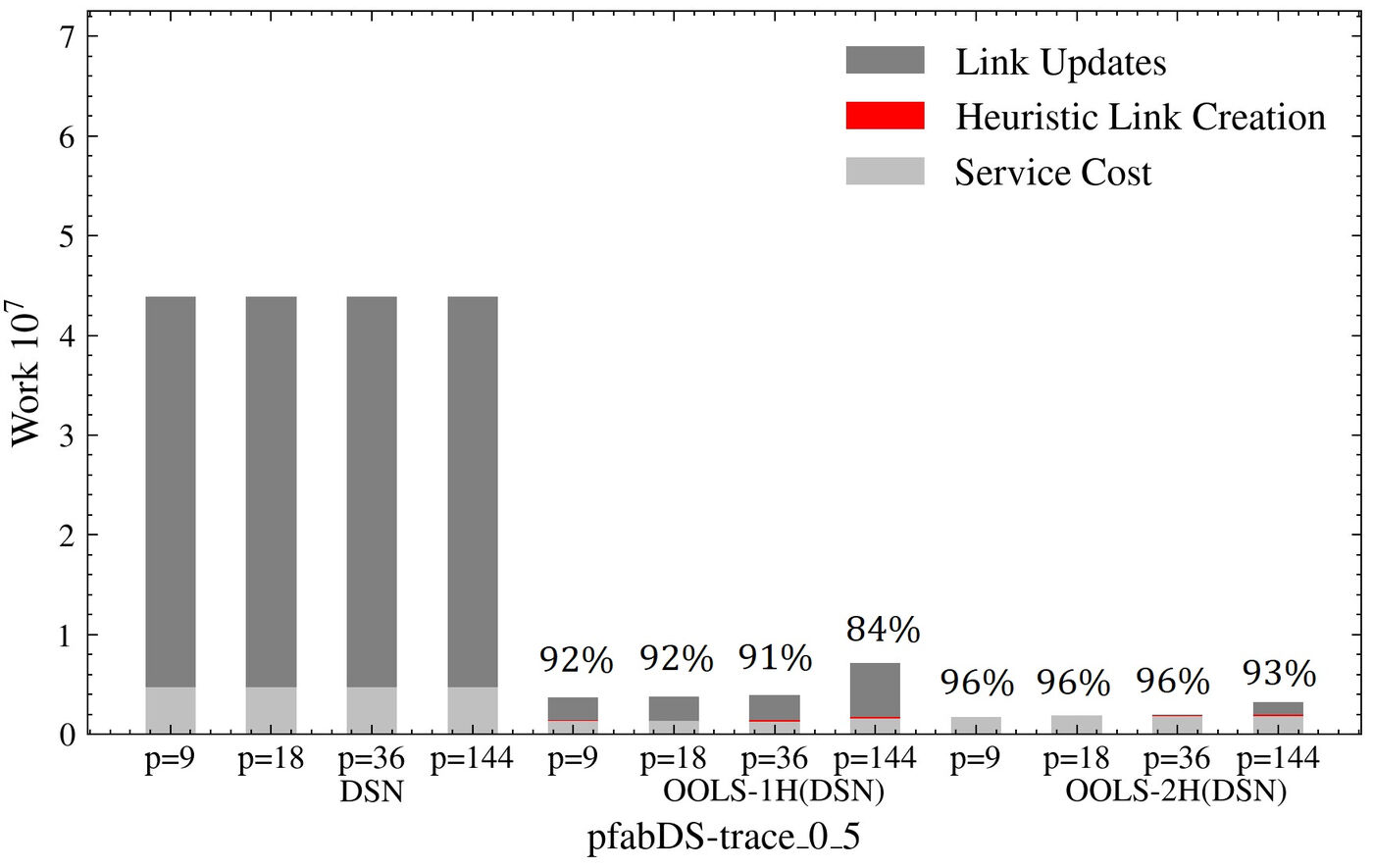}
    \caption{pFabric ($n=144, m=10^6$)}
    \label{fig:mu1_work_pfab_dsn}
\end{subfigure}
\hfill
\begin{subfigure}[b]{0.32\textwidth}
    \includegraphics[width=\textwidth]{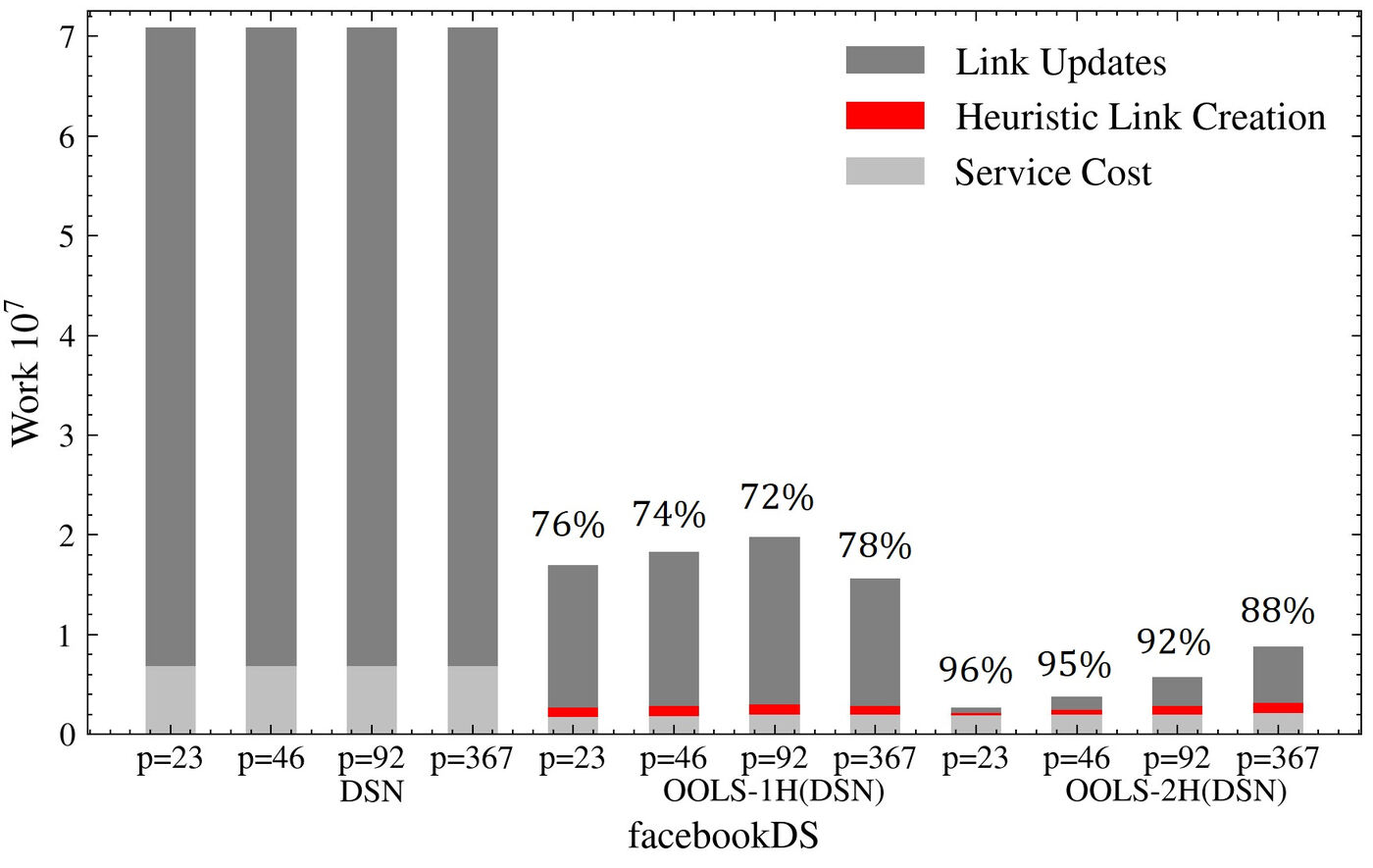}
    \caption{Facebook ($n=367,m=10^6$)}
    \label{fig:mu1_work_facebook_dsn}
\end{subfigure}
\hfill
\begin{subfigure}[b]{0.32\textwidth}
\includegraphics[width=\textwidth]{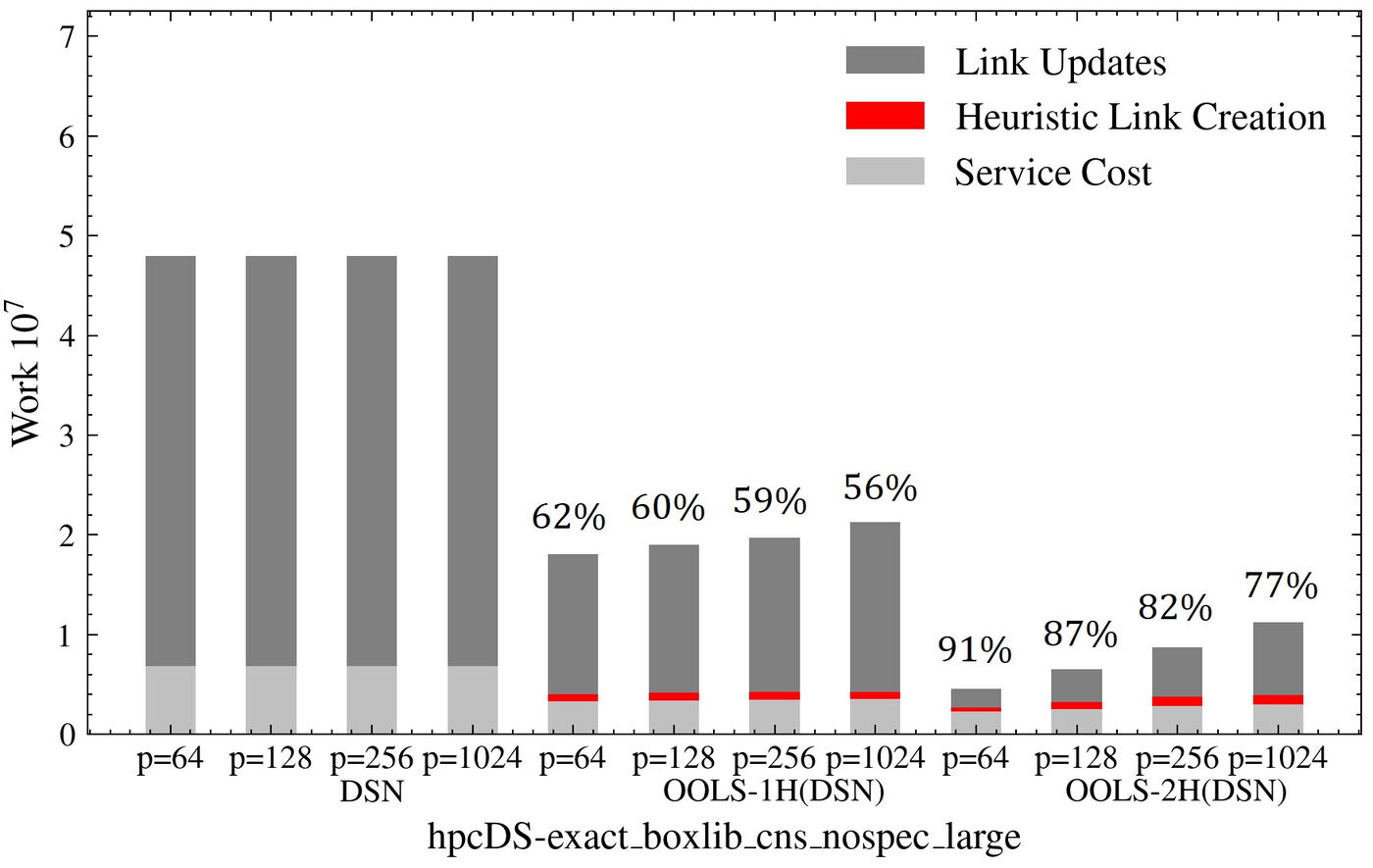}
\caption{HPC ($n=1024, m=10^6$)}
\label{fig:mu1_work_hpc_dsn}
\end{subfigure}
\hfill
\begin{subfigure}[b]{0.32\textwidth}
\includegraphics[width=\textwidth]{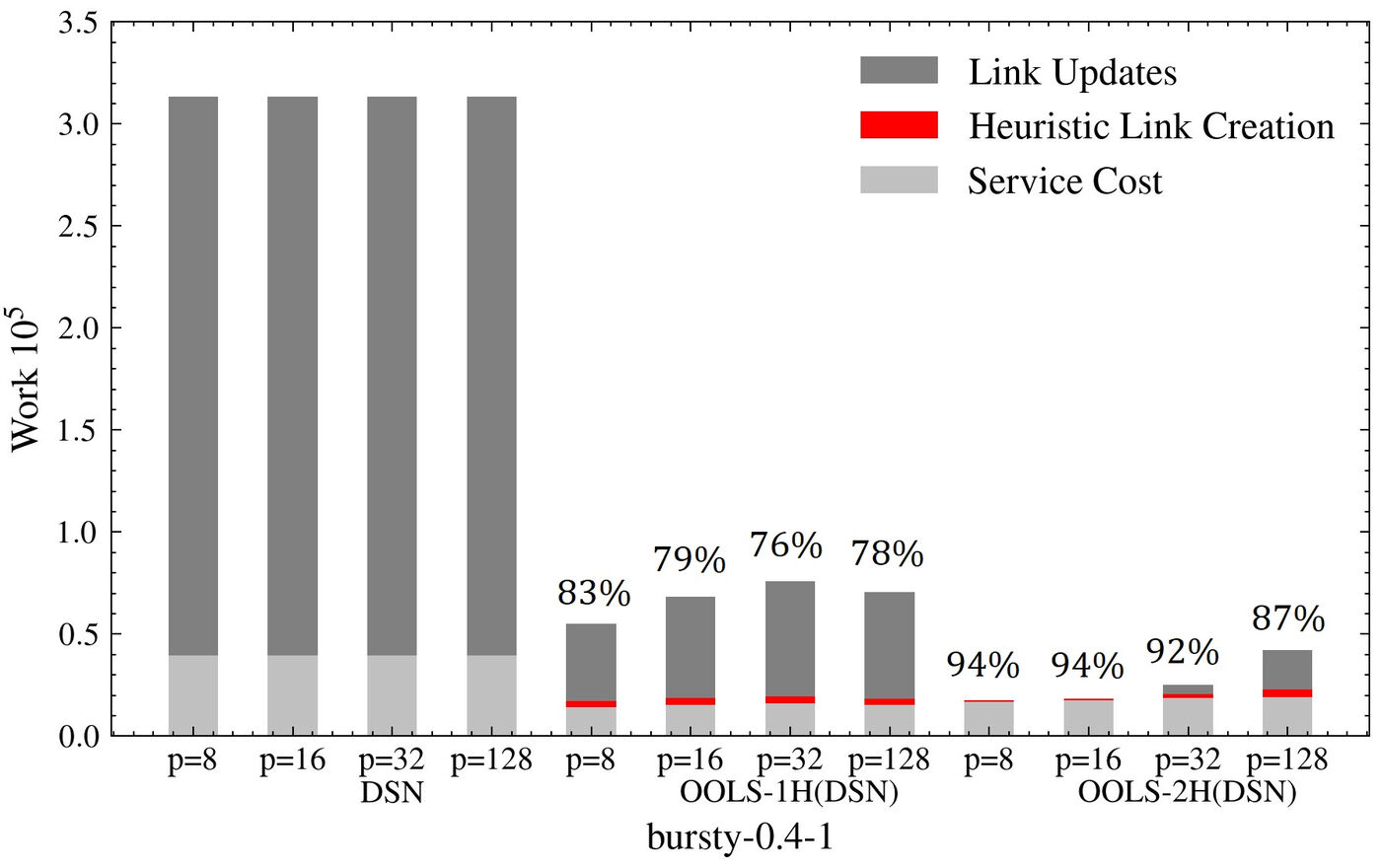}
\caption{Bursty ($n=128,p=8, m=10^4$)}
\label{fig:mu1_work_bursty_dsn}
\end{subfigure}
\hfill
\begin{subfigure}[b]{0.32\textwidth}
\includegraphics[width=\textwidth]{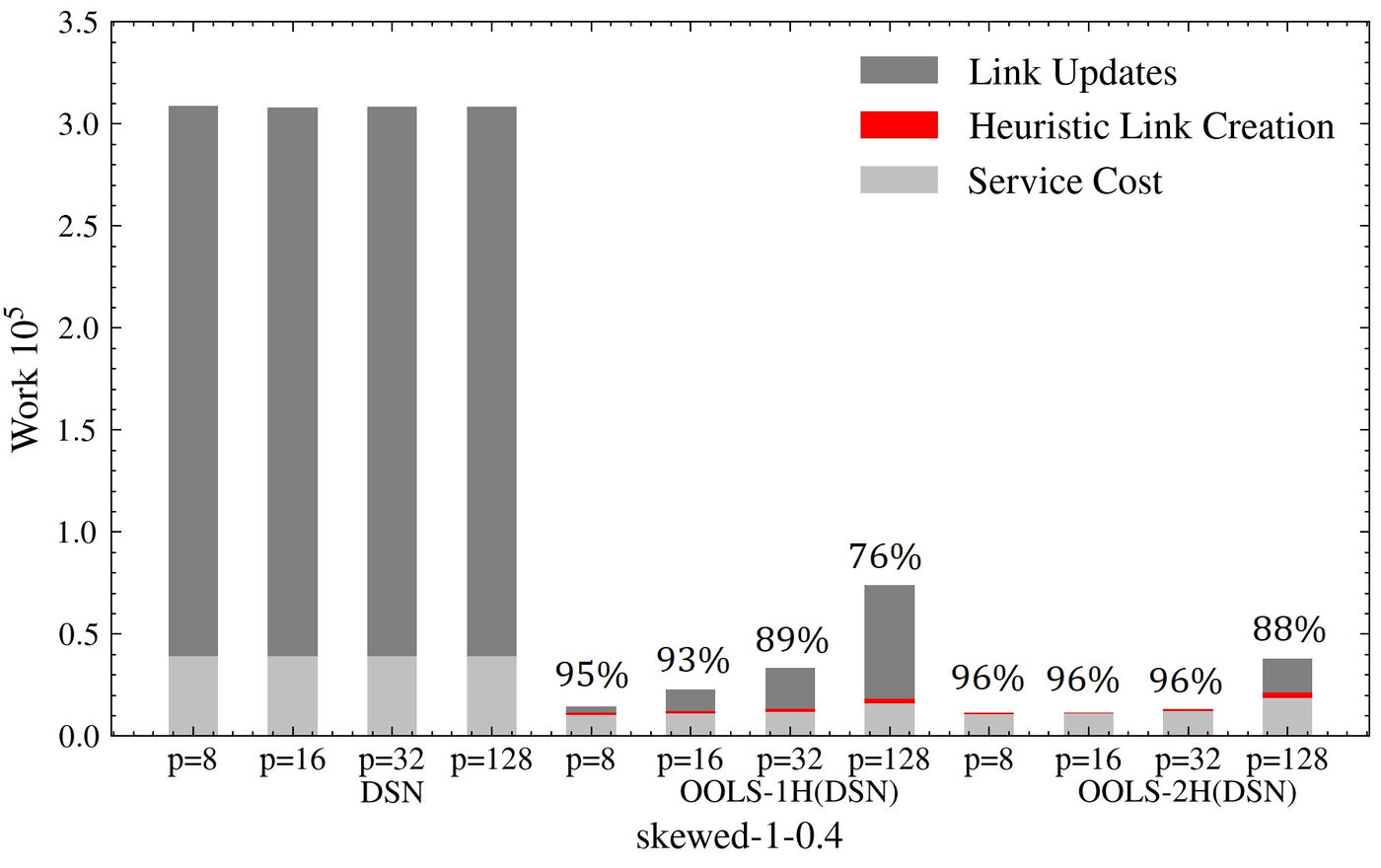}
\caption{Skewed ($n=128, m=10^4$)}
\label{fig:mu1_work_skewed_dsn}
\end{subfigure}
\hfill
\begin{subfigure}[b]{0.32\textwidth}
\includegraphics[width=\textwidth]{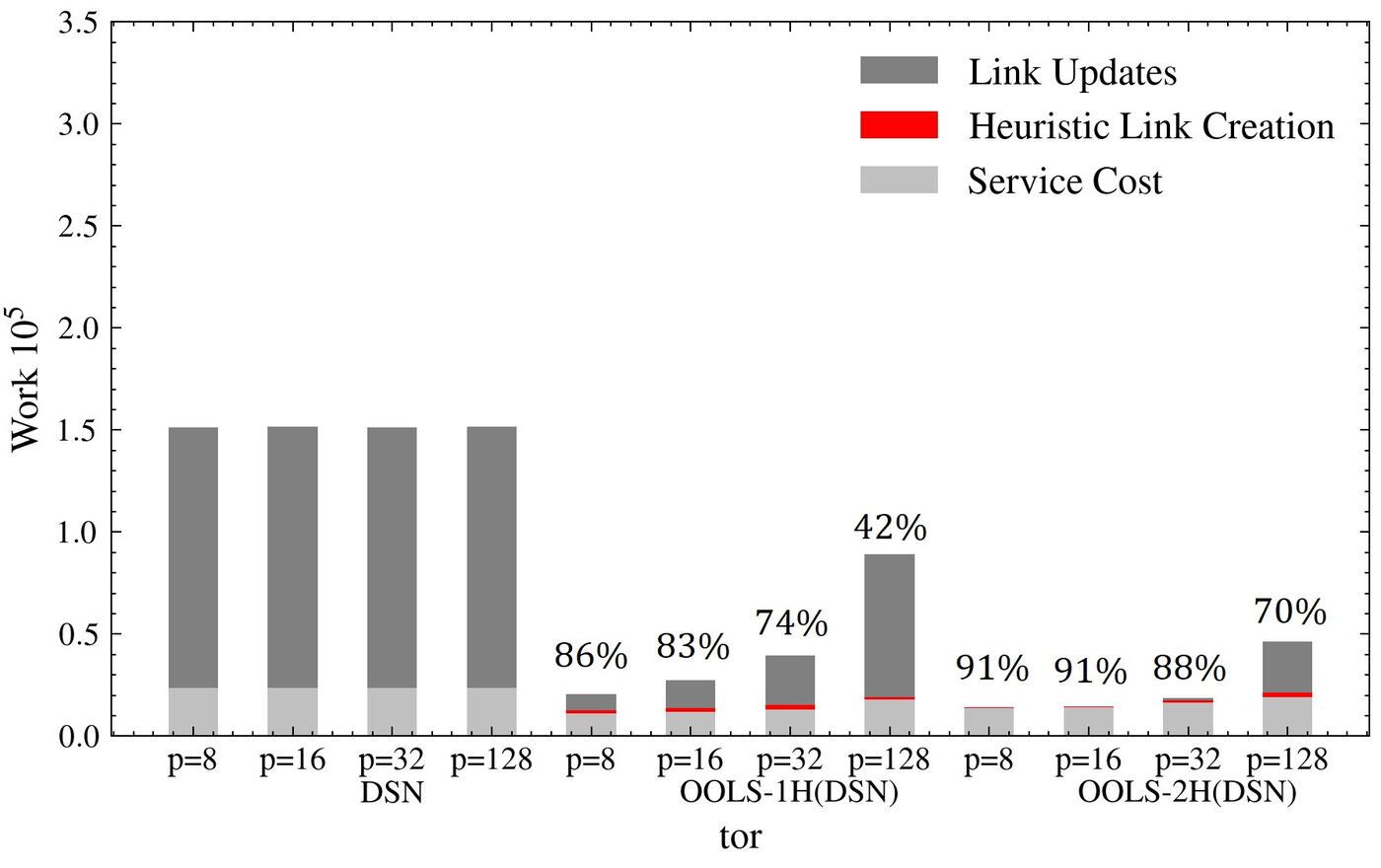}
\caption{ProjecToR ($n=128, m=10^4$)}
\label{fig:mu1_work_tor_dsn}
\end{subfigure}
\caption{Total work over all datasets (DSN-based SANs, var. switch size $p$). The \% numbers are relative to the cost of DSN.}
\label{fig:mu1_work_all_dsn}
\end{figure*}

In particular, for $p=\lceil{n}/{16}\rceil$, the makespan reductions achieved by \textit{one-hop} and \textit{two-hop} OOLS, resp., were: $81\%$ and $81\%$ in pFabric (Fig. \ref{fig:mu1_work_pfab_dsn}), $70\%$ and $82\%$ in Facebook (Fig. \ref{fig:mu1_work_facebook_dsn}), $40\%$ and $85\%$ in HPC (Fig. \ref{fig:mu1_work_hpc_dsn}), $75\%$ and $79\%$ in Bursty (Fig. \ref{fig:mu1_work_bursty_dsn}), $76\%$ and $76\%$ in Skewed (Fig. \ref{fig:mu1_work_skewed_dsn}), and $61\%$ and $66\%$ in ProjecToR (Fig. \ref{fig:mu1_work_tor_dsn}). Already in one-hop version, OOLS performed better in pFabric. In the same workload, OOLS maintained the same performance in the two-hop version. This may indicate that OOLS favors primarily temporal locality, but also higher values of $n$ (which distinguishes pFabric from Bursty) to reduce makespan.

\subsection{Communication Cost (Total Work)}
\label{subsec:results_work}

In Figure \ref{fig:mu1_work_all_dsn}, we analyze the total communication cost (Eq. (\ref{eq:total_cost})), partitioned into \textit{service cost} (Eq. (\ref{eq:srv_cost})), \textit{link updates} (Eq. (\ref{eq:link_cost})), and, additionally, \textit{opportunistic links} (created by OOLS), in all scenarios from Figure \ref{fig:mu1_work_all_dsn}, with the addition of a variable switch size $p \in [\lceil{n}/{16}\rceil, \lceil{n}/{8}\rceil, \lceil{n}/{4}\rceil, n]$. Moreover, we included the results for the \textit{one-hop} OOLS (OOLS-1H(DSN)) and excluded the CBN baseline. 

First, note that the baseline (DSN) is invariant to the value of $p$, as expected. In all datasets, OOLS decreases the total communication cost as it does with the makespan, especially with a lower $p$ value, given a fixed $n$. In particular, for $p= \left\lceil {n}/{16}\right\rceil$, two-hop OOLS reduces the total cost by $96\%$, $96\%$, $91\%$, $94\%$, $96\%$, $91\%$, for the pFabric (Fig. \ref{fig:mu1_work_pfab_dsn}), Facebook (Fig. \ref{fig:mu1_work_facebook_dsn}), HPC (Fig. \ref{fig:mu1_work_hpc_dsn}), Bursty (Fig. \ref{fig:mu1_work_bursty_dsn}), Skewed (Fig. \ref{fig:mu1_work_skewed_dsn}), and ProjecTor (Fig. \ref{fig:mu1_work_tor_dsn}) workloads, respectively. For $p= n$, these values become $93\%$, $88\%$, $77\%$, $87\%$, $88\%$, $70\%$, respectively.

Comparing \textit{one-hop} and \textit{two-hop} OOLS, the latter outperforms the former in all scenarios, also in total work reduction, confirming that after all one-hop link opportunities have been used up, two-hop links remain available for OOLS. In particular, when $p=\lceil{n}/{16}\rceil$, the cost reductions achieved by \textit{one-hop} and \textit{two-hop} OOLS, resp., were: $92\%$ and $96\%$ in pFabric (Fig. \ref{fig:mu1_work_pfab_dsn}), $76\%$ and $96\%$ in Facebook (Fig. \ref{fig:mu1_work_facebook_dsn}), $62\%$ and $91\%$ in HPC (Fig. \ref{fig:mu1_work_hpc_dsn}), $83\%$ and $94\%$ in Bursty (Fig. \ref{fig:mu1_work_bursty_dsn}), $95\%$ and $96\%$ in Skewed (Fig. \ref{fig:mu1_work_skewed_dsn}), and $86\%$ and $91\%$ in ProjecToR (Fig. \ref{fig:mu1_work_tor_dsn}).

\begin{figure*}[t]
\centering	
\begin{subfigure}[b]{0.32\textwidth}
\includegraphics[width=\textwidth]{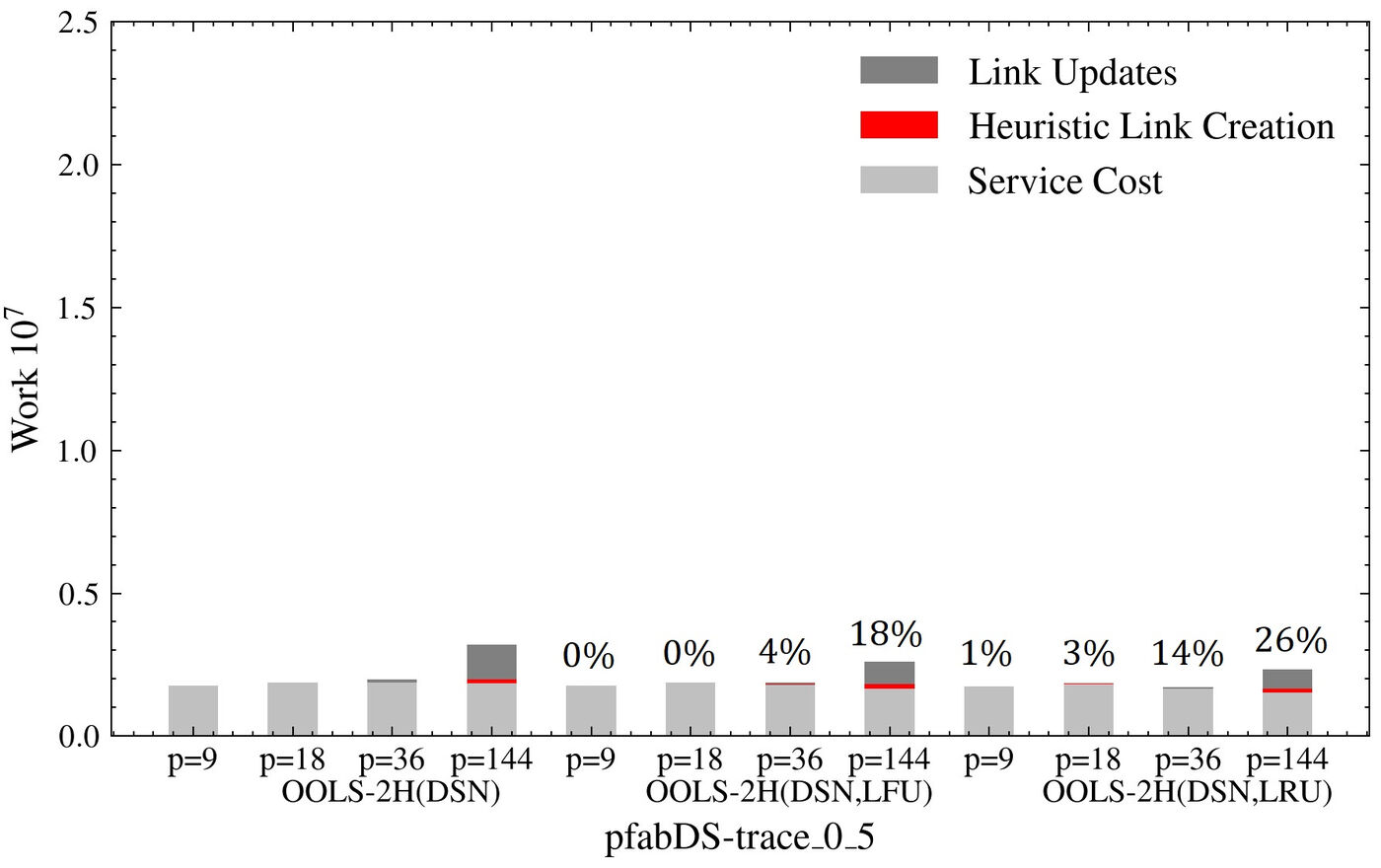}
\caption{pFabric ($n=144, m=10^6$)}
\label{fig:mu1_work_2h_pfab_dsn}
\end{subfigure}
\hfill
\begin{subfigure}[b]{0.32\textwidth}
\includegraphics[width=\textwidth]{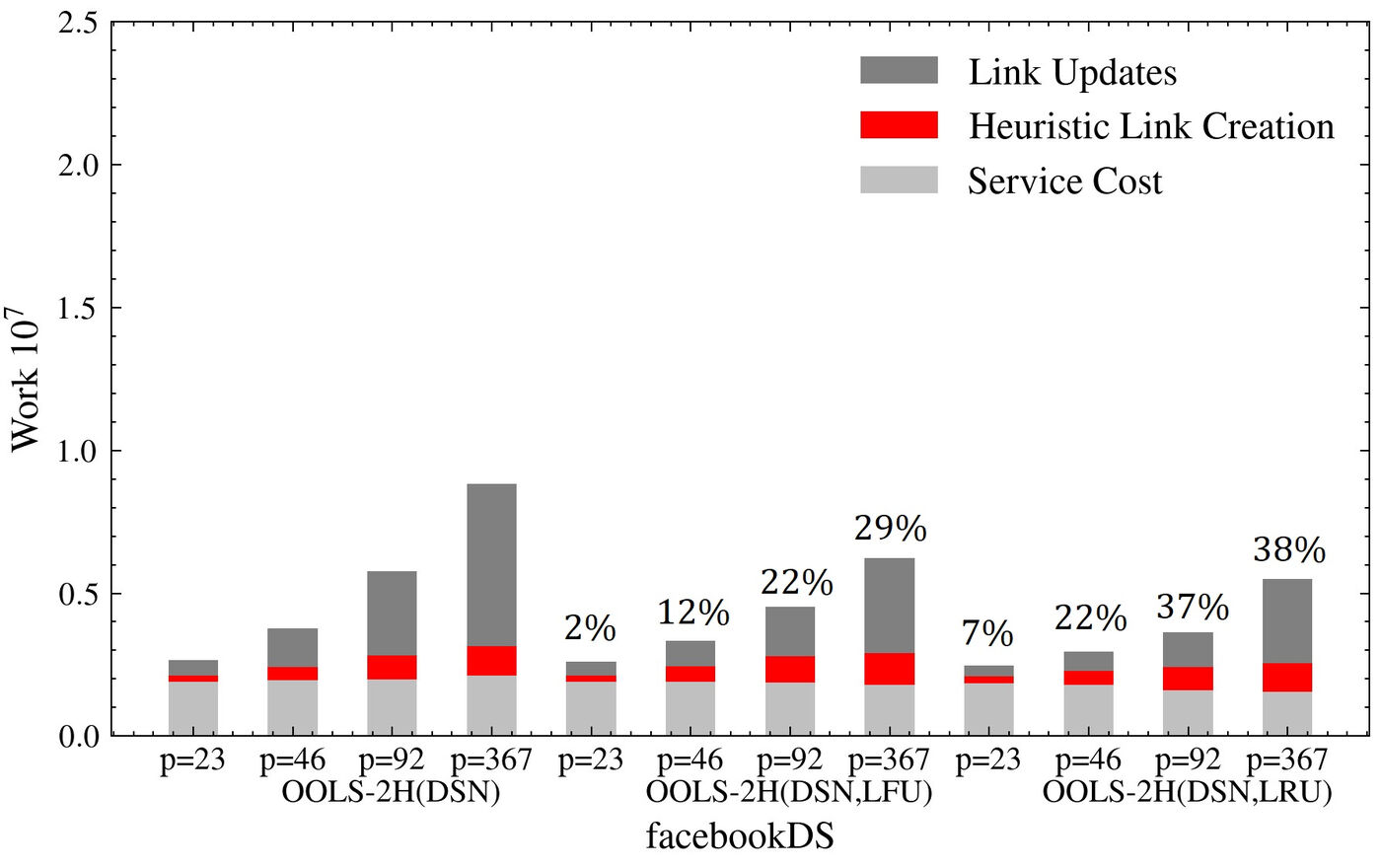}
\caption{Facebook ($n=367, m=10^6$)}
\label{fig:mu1_work_2h_facebook_dsn}
\end{subfigure}
\hfill
\begin{subfigure}[b]{0.32\textwidth}
\includegraphics[width=\textwidth]{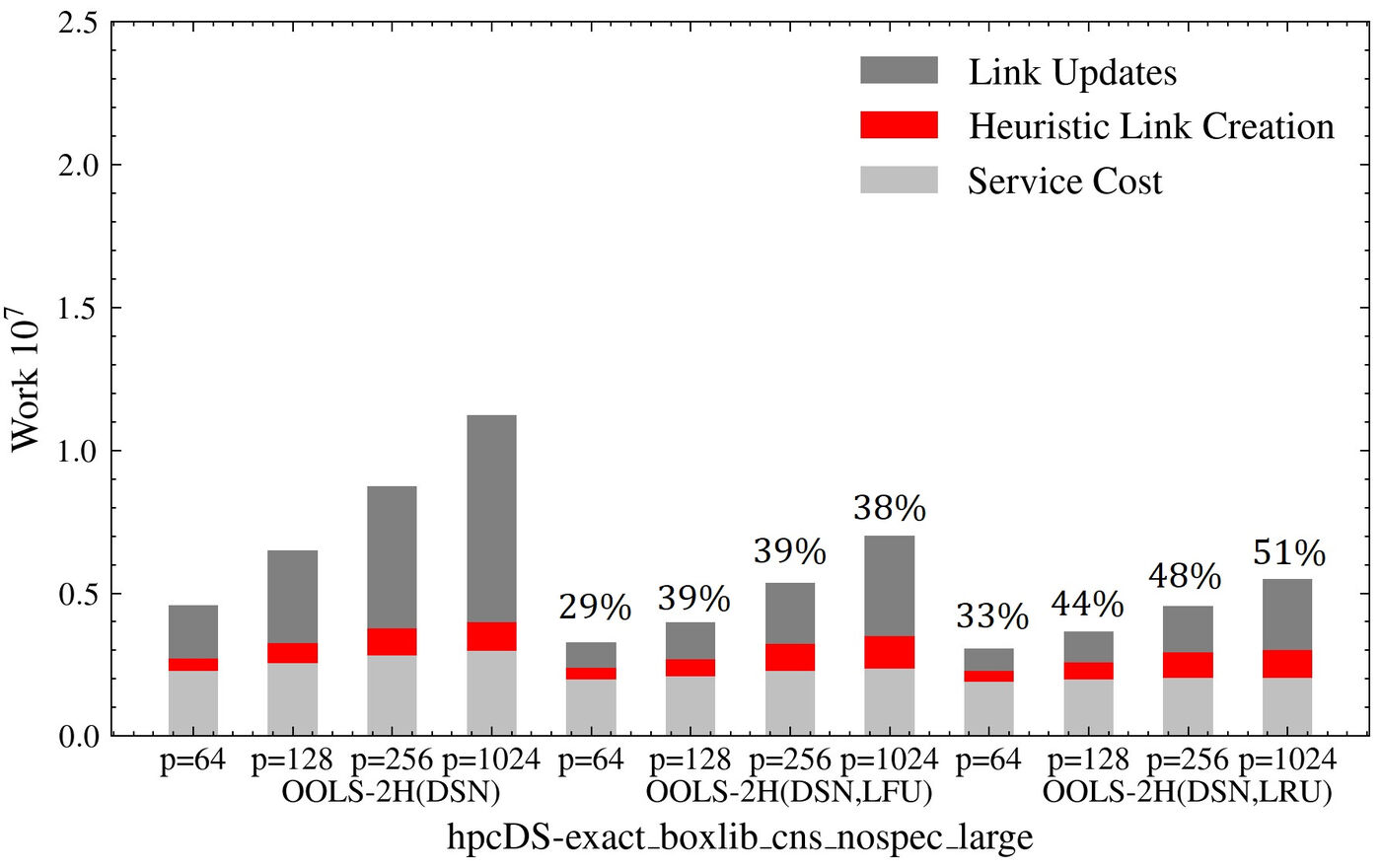}
\caption{HPC ($n=1024, m=10^6$)}
\label{fig:mu1_work_2h_hpc_dsn}
\end{subfigure}
\caption{OOLS with Link Replacement (LR): Total work (variable $p$). The \% numbers are w.r.t. the cost of OOLS without LR.}
\label{fig:mu1_work_all_2h_dsn}
\end{figure*}

\subsection{Link (cache) Replacement}
\label{subsec:results_cache}

In Figure \ref{fig:mu1_work_all_2h_dsn}, we analyze the impact of cache replacement strategies (LFU and LRU) on the total communication cost of OOLS (two-hop, DSN-based). We focus on workloads with variable network size (pFabric $(n=144)$, Fig. \ref{fig:mu1_work_2h_pfab_dsn}, Facebook $(n=367)$, Fig. \ref{fig:mu1_work_2h_facebook_dsn}, and HPC $(n=1024)$, Fig. \ref{fig:mu1_work_2h_hpc_dsn}) and scenarios with variable switch size ($p \in [\lceil {n}/{16}\rceil, \lceil {n}/{8}\rceil, \lceil{n}/{4}\rceil, n]$).

Firstly, note that because DSN is a SAN algorithm that prioritizes rotations over message forwarding, its \textit{service cost} generally approximates $m$, i.e. a constant close to a lower bound of one hop per message. Therefore, even though OOLS is designed to reduce both the \textit{service cost} and the \textit{link update cost}, in the case of DSN, OOLS affects almost exclusively the \textit{link update cost}.

As a result, the gain obtained by OOLS and, in particular, by cache replacement heuristics, is very small in scenarios where the number of rotations is close to zero. This is what we observe in the pFabric scenario (Fig. \ref{fig:mu1_work_2h_pfab_dsn}): the \textit{link update} portion of the total communication cost of OOLS is close to zero for all values of $p$ except $p=n$ and, therefore, the total communication cost is almost a constant, with or without cache replacement.

In the scenarios with higher \textit{link update cost} (Fig. \ref{fig:mu1_work_2h_facebook_dsn} and Fig. \ref{fig:mu1_work_2h_hpc_dsn}), we observe that link (cache) replacement significantly reduces the total communication cost of OOLS, which increases with the switch size ($p$). E.g., even when $p=n$, \texttt{OOLS-2H(DSN, LRU)} reduced the total work by $51\%$ in HPC, and by $38\%$ in Facebook, relative to \texttt{OOLS-2H} without link replacement. 

Finally, we observe that LRU slightly outperforms LFU, especially in workloads with higher temporal locality, like pFabric (Fig. \ref{fig:mu1_work_2h_pfab_dsn}) and HPC (Fig. \ref{fig:mu1_work_2h_hpc_dsn}).

\begin{figure*}[h]
\centering
\begin{subfigure}[b]{0.32\textwidth}
\includegraphics[width=\textwidth]{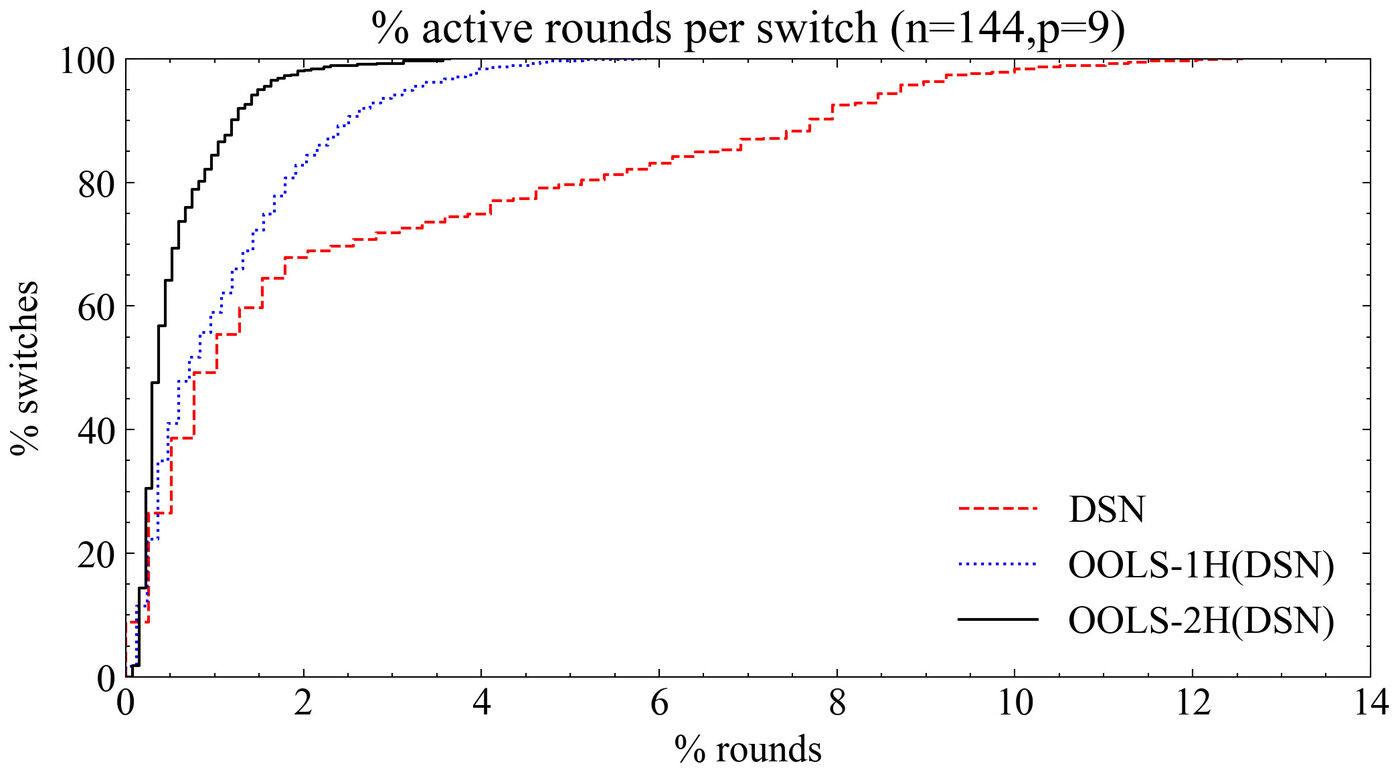}
\caption{pFabric ($n=144, p=9, m=10^6$)}
\label{fig:mu1_active_switches_pfab_dsn}
\end{subfigure}
\hfill
\begin{subfigure}[b]{0.32\textwidth}
\includegraphics[width=\textwidth]{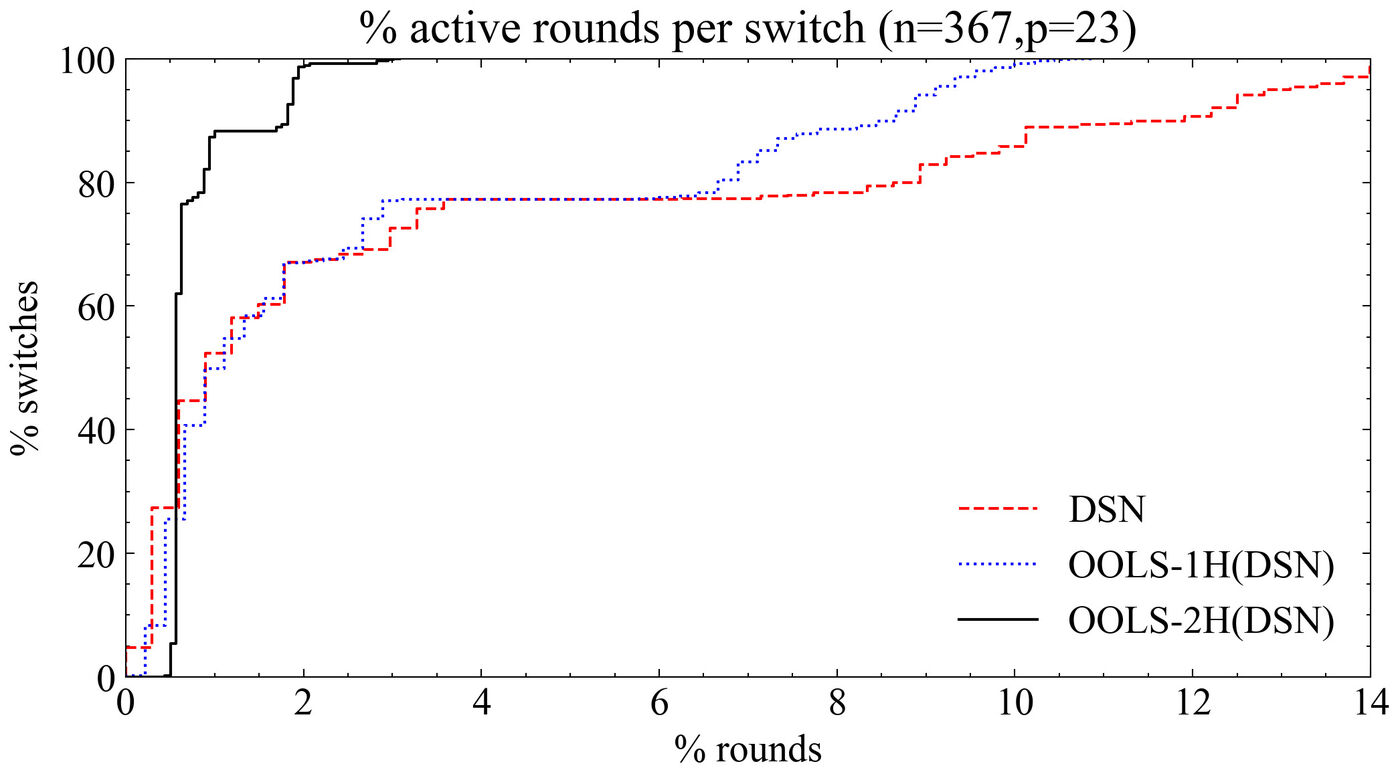}
\caption{Facebook ($n=367, p=23, m=10^6$)}
\label{fig:mu1_active_switches_facebook_dsn}
\end{subfigure}
\hfill
\begin{subfigure}[b]{0.32\textwidth}
\includegraphics[width=\textwidth]{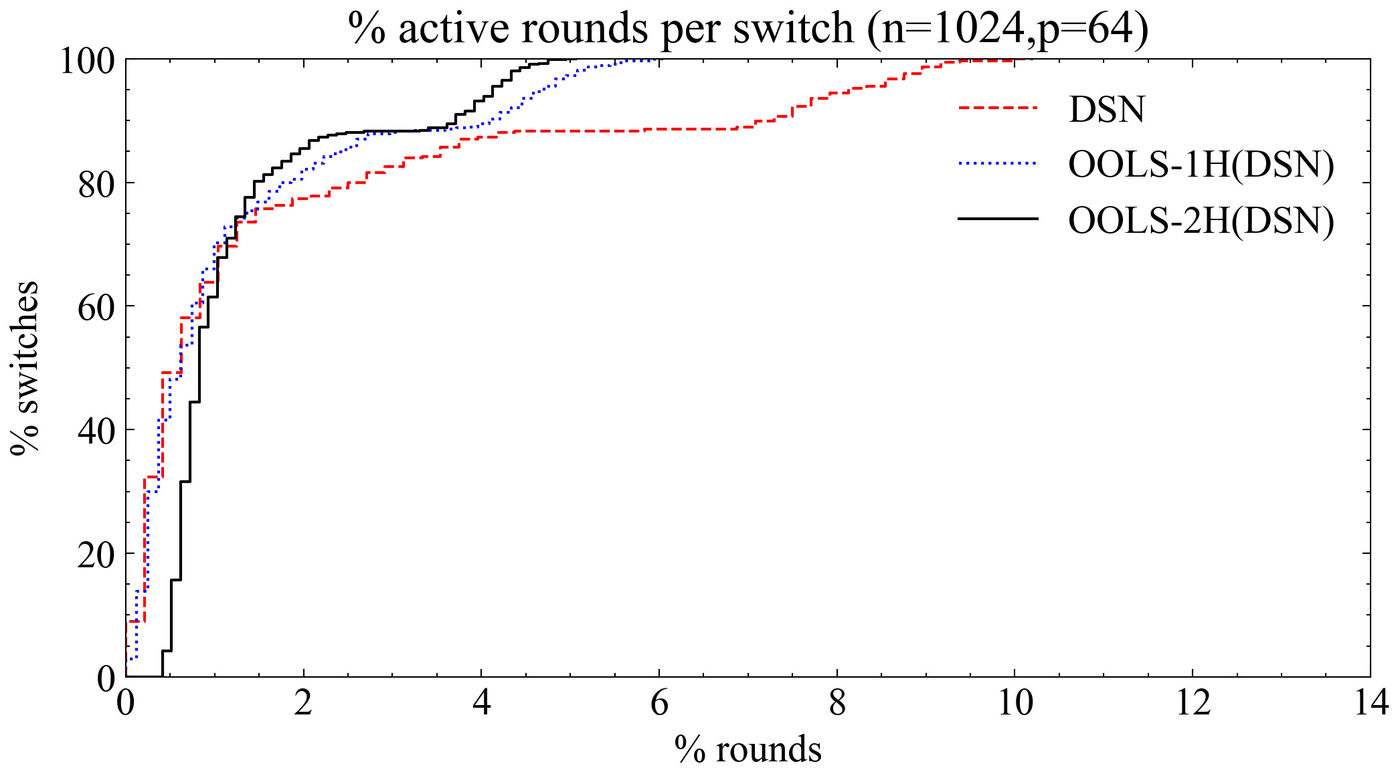}
\caption{HPC ($n=1024, p=64, m=10^6$)}
\label{fig:mu1_active_switches_hpc_dsn}
\end{subfigure}
\hfill
\begin{subfigure}[b]{0.32\textwidth}
\includegraphics[width=\textwidth]{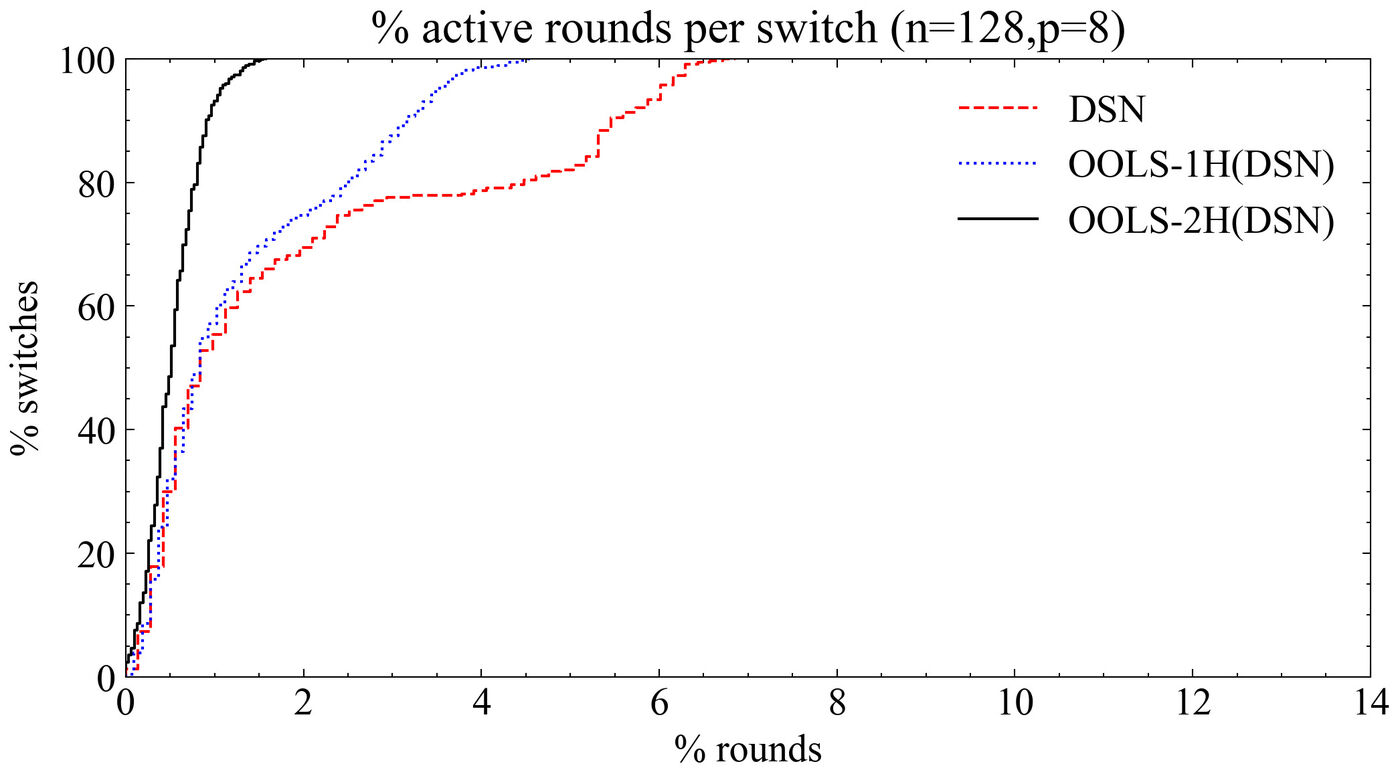}
\caption{Bursty ($n=128,p=8, m=10^4$)}
\label{fig:mu1_active_switches_bursty_dsn}
\end{subfigure}
\hfill
\begin{subfigure}[b]{0.32\textwidth}
\includegraphics[width=\textwidth]{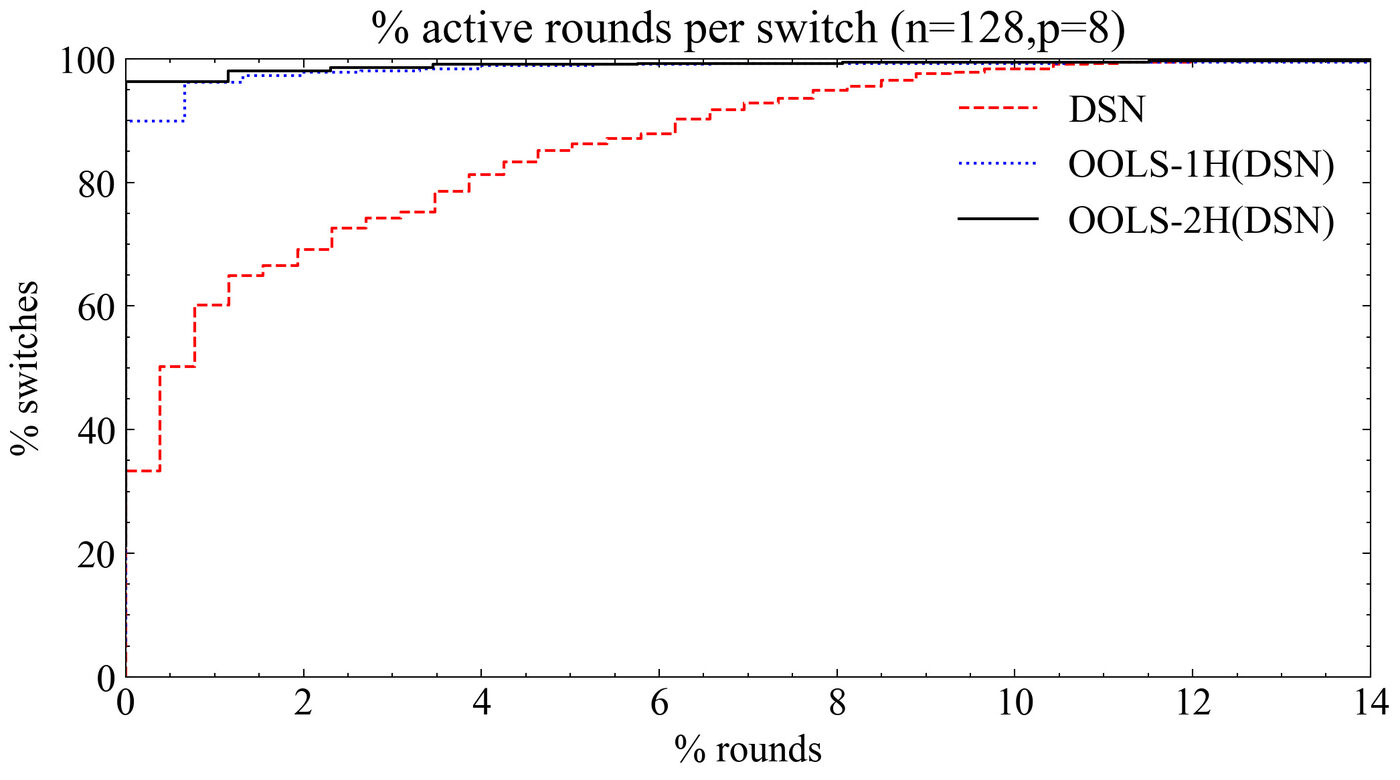}
\caption{Skewed ($n=128,p=8, m=10^4$)}
\label{fig:mu1_active_switches_skewed_dsn}
\end{subfigure}
\hfill
\begin{subfigure}[b]{0.32\textwidth}
\includegraphics[width=\textwidth]{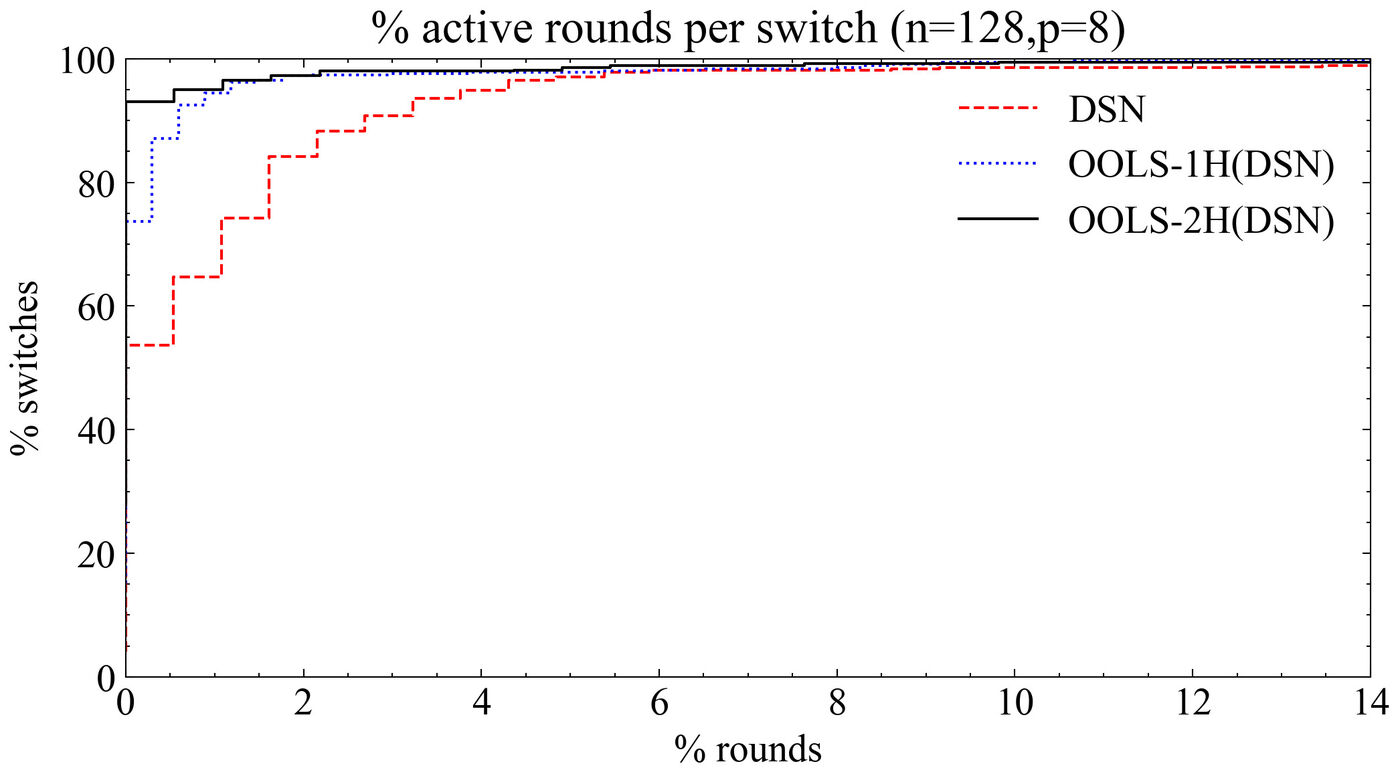}
\caption{ProjecToR ($n=128,p=8, m=10^4$)}
\label{fig:mu1_active_switches_tor_dsn}
\end{subfigure}
\caption{CDF \% active switches per round over all datasets (DSN-based SANs, $\lceil {n}/{p}\rceil = 16$).}
\label{fig:mu1_active_switches_all_dsn}
\end{figure*}

\subsection{OCS Activity Distribution (constant switch size)}
\label{subsec:results_cdf_constant_p}

In Figure \ref{fig:mu1_active_switches_all_dsn}, we show the CDF (Cumulative Distribution Function), over the set of all optical switches, of the percentage of rounds (relative to the total simulation time) in which at least one message has been routed through each switch. We compare the behavior of \textit{one-hop} OOLS, \textit{two-hop} OOLS, both running in combination with DSN, to the DSN baseline (without opportunistic link scheduling). Note that a rapidly rising CDF curve points to sparser activity bursts, while a slower-rising curve indicates a more even distribution of OCS activity in time.

We observe that, in all simulated scenarios, the fastest-rising CDF curve is that of the \textit{two-hop} OOLS, and the slowest-rising curve is that of the baseline (DSN), which reveals that optical switches are less active over time in OOLS than in DSN. This can be explained by the fact that OOLS reduces the number of hops needed to deliver some messages, which reduces the total communication cost, resulting in lower switch activity over time.

In workloads with very high spatial locality, like Skewed (Fig. \ref{fig:mu1_active_switches_skewed_dsn}) and ProjecTor (Fig. \ref{fig:mu1_active_switches_tor_dsn}), the \textit{two-hop} OOLS exhibited the highest OCS inactivity. In the former scenario, approximately $95\%$ of switches were active $0\%$ of the rounds, i.e., completely idle. In the latter, almost $92\%$ of the switches were \textit{inactive} $100\%$ of the time. This is due to the small OCS size ($p= \lceil{n}/{16}\rceil$), which results in a greater number of OCS switches used by OpticNet, whereas the demand is concentrated at an extremely small subset of nodes. Note that in all the other workloads, $0\%$ of switches remained inactive $100\%$ of the time.

In workloads with \textit{high spatial locality and high temporal locality}, like HPC (Fig. \ref{fig:mu1_active_switches_hpc_dsn}), (two-hop) OOLS exhibited the lowest OCS inactivity: approx. $95\%$ of switches were active at least $4\%$ of the simulation rounds. This is due to the demand, which is spread over a larger subset of nodes, the SAN topology, which is more dynamic, and the greater total reconfiguration cost (number of \textit{link updates}).

\begin{figure*}[h]
\centering	
\begin{subfigure}[b]{0.32\textwidth}
\includegraphics[width=\textwidth]{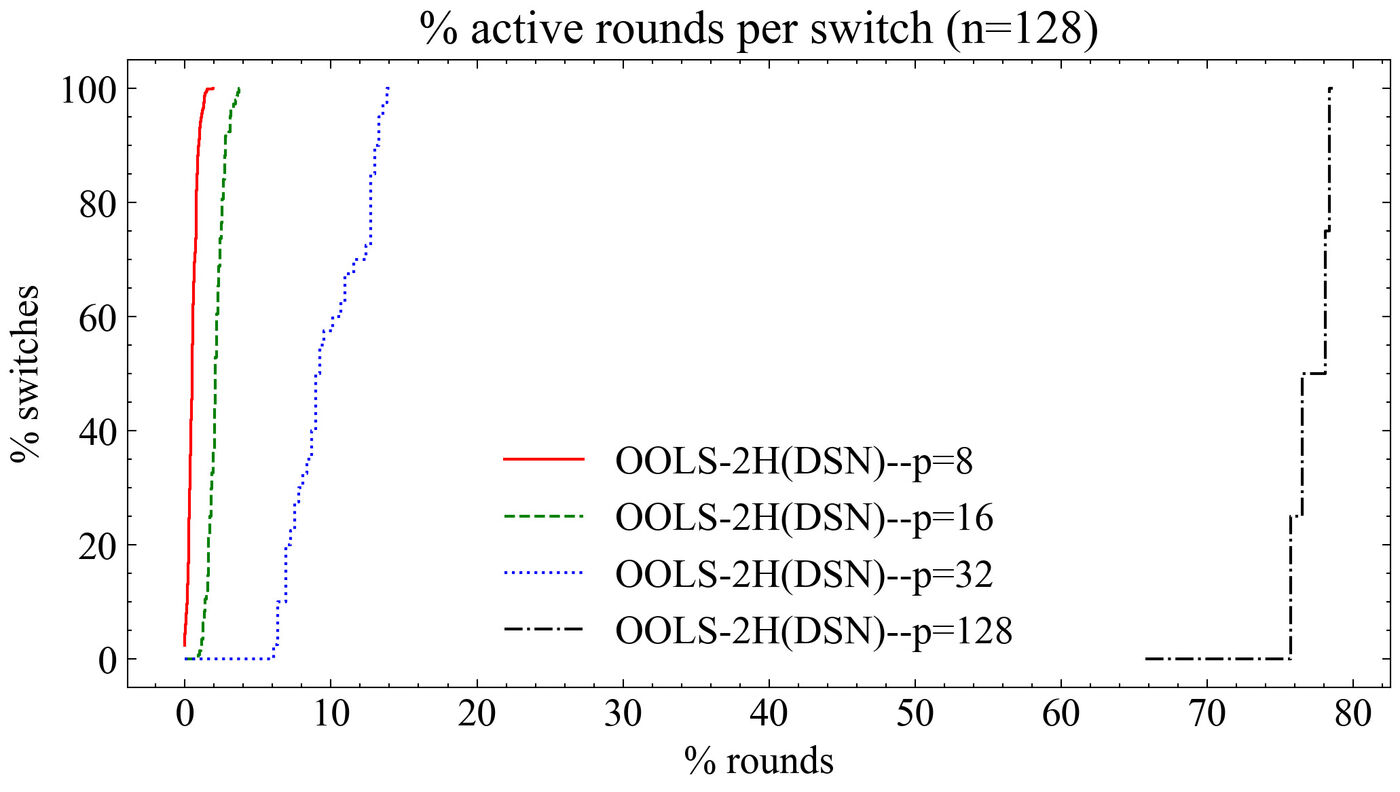}
\caption{Bursty ($n=128,p=8, m=10^4$)}
\label{fig:active_switches_2h_bursty_dsn}
\end{subfigure}
\hfill
\begin{subfigure}[b]{0.32\textwidth}
\includegraphics[width=\textwidth]{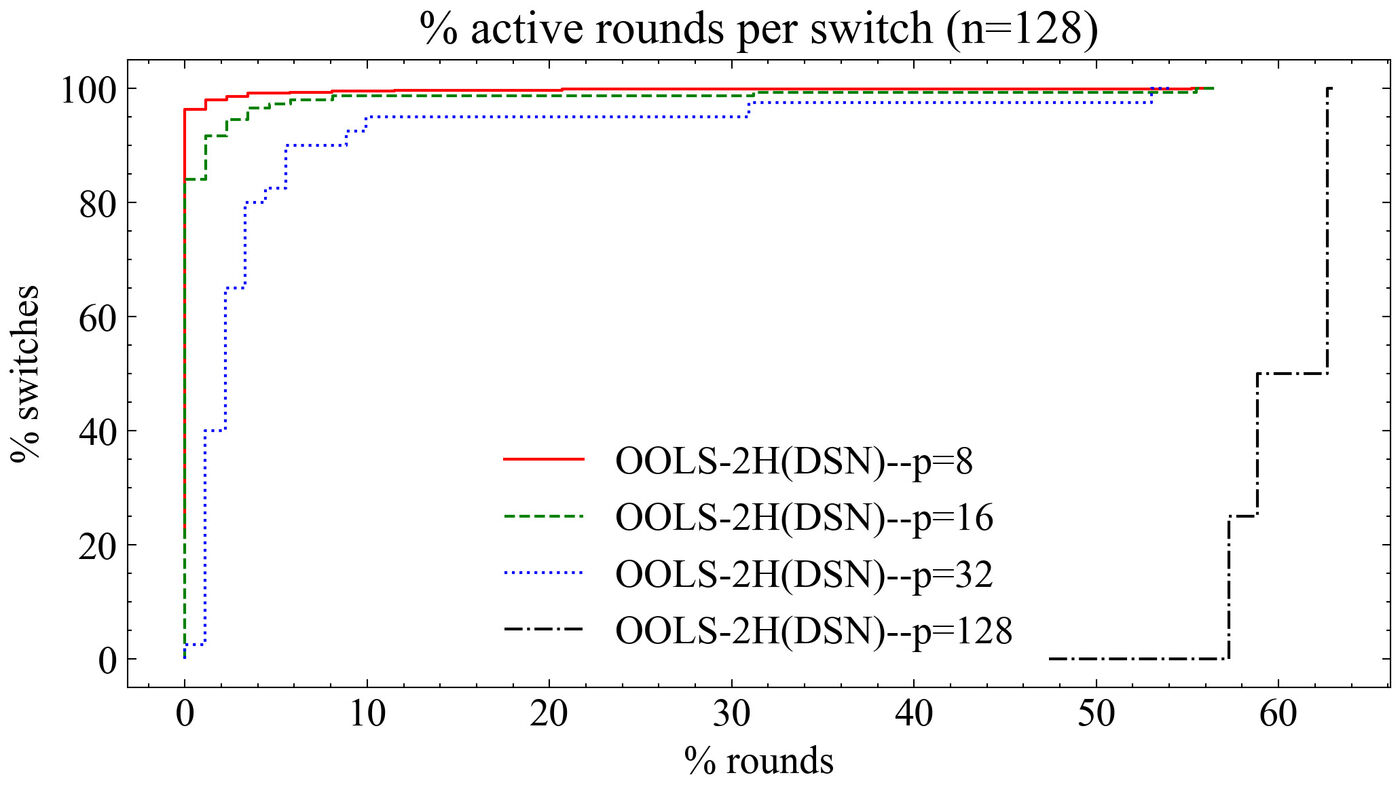}
\caption{Skewed ($n=128,p=8, m=10^4$)}    \label{fig:active_switches_2h_skewed_dsn}
\end{subfigure}
\hfill
\begin{subfigure}[b]{0.32\textwidth}
\includegraphics[width=\textwidth]{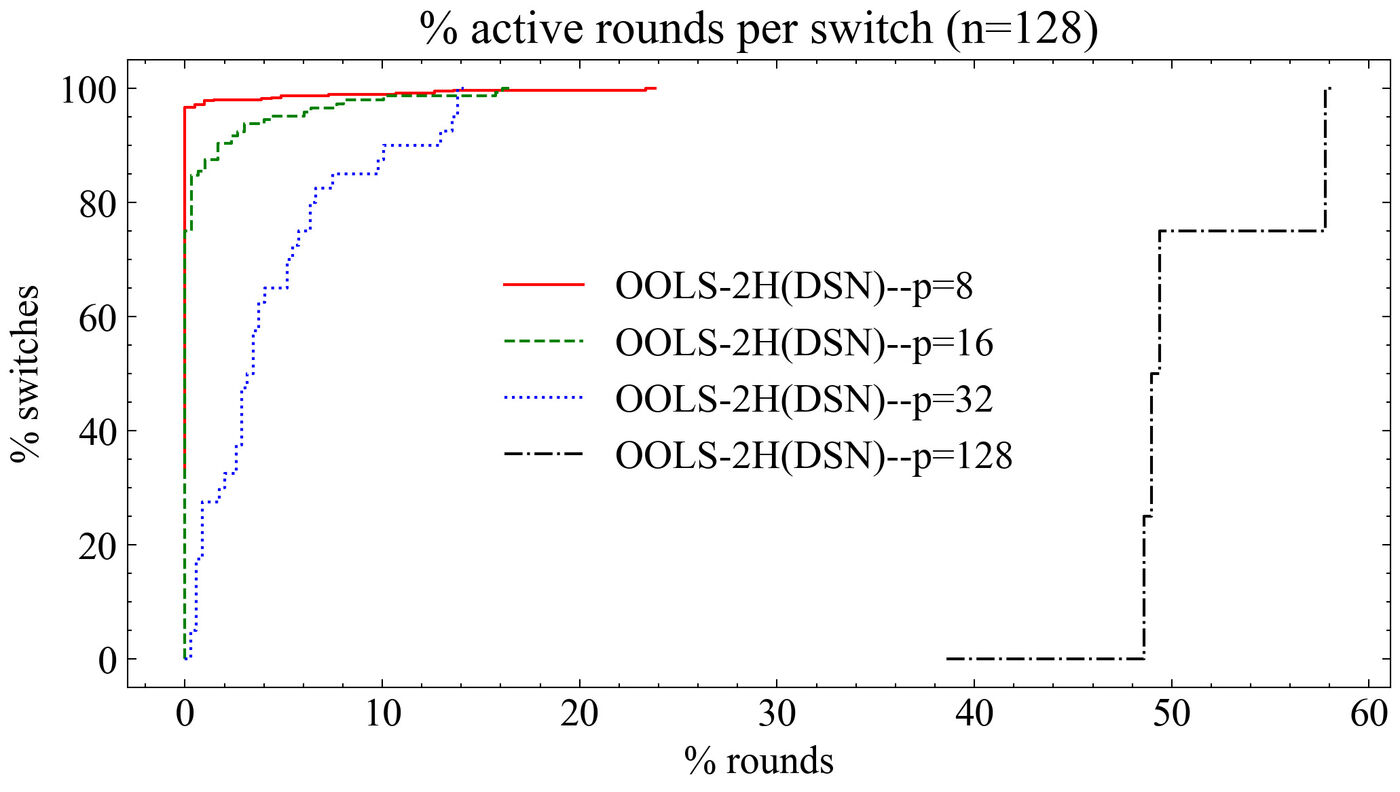}
\caption{ProjecToR ($n=128,p=8, m=10^4$)}
\label{fig:active_switches_2h_tor_dsn}
\end{subfigure}
\hfill
\caption{OOLS-2H(DSN): CDF \% active switches per round (variable switch size $p$).}
\label{fig:active_switches_2h_all_dsn}
\end{figure*}

\subsection{OCS Activity Distribution (variable switch size)}
\label{subsec:results_cdf_var_p}

We now analyze the behavior of the \textit{two-hop} OOLS (running in combination with DSN) in scenarios with increasing OCS size (number of ports) $p\in [8, 16, 32, 128]$, in three scenarios with network size $n=128$: Bursty, Skewed, and ProjecToR. 
Figure \ref{fig:active_switches_2h_all_dsn}
depicts the CDF (over the set $SW$) of the \% of active rounds per switch.

We observe that, in all scenarios (Fig. \ref{fig:active_switches_2h_bursty_dsn}, Fig. \ref{fig:active_switches_2h_skewed_dsn}, Fig. \ref{fig:active_switches_2h_tor_dsn}), 
the CDF curve is shifted to the right with increasing values of $p$, i.e., a greater number of switches are activated in a greater number of simulation rounds, and a greater \% of ports per switch are activated at least once during the simulation. E.g., in the Skewed workload, for $p=8$, only $4\%$ of switches are activated at least once during the entire simulation (Fig. \ref{fig:active_switches_2h_skewed_dsn}, red line).
On the other hand, when $p=128$, all $100\%$ (4 out of 4) of switches are activated $\geq 50\%$ of the time (Fig. \ref{fig:active_switches_2h_skewed_dsn}, black line).
This is because, when $p$ \textit{increases}, even though the demand sequence remains the same, it is spread over fewer available OCS ports, resulting in a higher switch activation ratio.

\subsection{Final remarks}
\label{subsec:results_conclusion}

In all workloads, OOLS significantly increased the \textit{concurrency} of the system, relative to the baseline SAN algorithms (DiSplayNet and CBNet). 
E.g., for $p= \left\lceil {n}/{16}\right\rceil$, the two-hop OOLS achieved a reduction in \textit{makespan} of at least $70\%$ in all workloads, relative to DSN.
The reduction in communication cost was even greater than in makespan. The two-hop OOLS reduced the total cost by at least $91\%$ and $70\%$, for $p= \left\lceil {n}/{16}\right\rceil$ and  $p= n$, respectively. Comparing \textit{one-hop} and \textit{two-hop} OOLS, the latter outperforms the former in all scenarios. The performance of OOLS is further enhanced by the use of cache (link) replacement strategies (LRU or LFU), especially in scenarios with high reconfiguration costs and high temporal locality.

The simulation results pointed to a \textit{trade-off} between efficiency improvement and resource utilization, as a function of switch size $p$. On the one hand, when $p$ \textit{decreases}, OOLS achieves lower switch and port \% activation but greater gains in throughput and communication cost.
This is due to the fact that the number of idle OCS ports to be allocated opportunistically by OOLS is upper bounded by the total number of ports, given by $opp\leq p\times |SW| -n+1$, where $|SW|$ is the total number of optical switches in the OpticNet system (defined in Eq. (\ref{eq:numBSTswitches})), and $(n-1)$ is the minimum number of edges in any connected topology. For a BST network on $n$ nodes, 
$opp$ is inversely proportional to $p$:
$opp \leq p\cdot |SW|-n = p\left\lceil{n^2}/{p^2}\right\rceil + p\left\lceil{n}/{p}\right\rceil -n \leq \left\lceil {n^2}/{p}\right\rceil$.

On the other hand, when $p$ \textit{increases}, the demand is spread over fewer OCS ports, resulting in higher switch and port \% activation, but also lower gains in throughput and communication cost, relative to the baselines.

\section{Related Work}
\label{sec:related}

RDNs can be categorized based on whether topology reconfigurations are driven by traffic demand. Demand-oblivious designs \cite{rotornet, opera, sirius, 2022STOC_AMIR} operate according to fixed reconfiguration schedules, independent of workload characteristics, whereas demand-aware approaches 
\cite{2022Jupiter, projector, apocs21renets, splaynet, displaynet}
explicitly adapt connectivity in response to observed traffic.
Hybrid architectures combining both paradigms have also been explored~\cite{sigmetrics22cerberus}. 
Making decisions at sub-second timescales limits the ability to gather complete traffic information, suggesting the need for hybrid control designs that combine distributed, centralized, and possibly hierarchical coordination \cite{2022Jupiter}. 

So far, there seem too be no consensus in the community on which RDN design is optimal \cite{2025CACM}.
In \cite{2023TNSM}, the problem of Maximum Switching Capacity was formulated, as follows: given a set of $n$ terminal nodes and a set of identical  $p\times p$ switches, the objective is to maximize the total number of connections, while ensuring the strict non-blocking condition, i.e., that any point-to-point connection can be established. In \cite{2025TNSM3}, a so-called multiple-plane twisted-folded Clos network (MP-TF Clos) was proposed, to increase switching capacity, while relaxing the strict non-blocking requirement. This type of design incorporates intermediary switches between terminal node pairs, which requires a connection admission control mechanism.



\section{Conclusion}
\label{sec:conclusion}



In this work, we presented OOLS, comprised of a set of lightweight real-time routines, designed to leverage idle OCS resources in RDN architectures. Our work is based on the Scalable Matching Model (SMM) \cite{2024tccOpticnet}. 
SMM supports small OCS devices in exchange for a set of ($n$) larger (ToR) electrical packet switches (typically available for a lower per-port price than OCS hardware \cite{polatis}).
SMM guarantees strict non-blocking, uses a minimum number of OCSs to realize any (dynamic) bounded-degree network topology, and allows executing existing SAN algorithms on top of it while preserving the performance (amortized communication cost) guarantees of the latter.
In our experiments, OOLS significantly improved the performance of SAN algorithms, lowering communication cost and boosting throughput.

We conclude by noting that RDNs are a promising but still experimental concept. 
Many questions remain open regarding the effects of dynamic topologies on higher protocol layers \cite{2024TNSM, 2022TNSM}, static and dynamic optimality of SAN algorithms \cite{apocs21renets, 2019SODA, Tango_SICOMP}, and the optimal trade-off between maximizing throughput while minimizing latency \cite{2022STOC_AMIR, 2025TNSM2}. 






\bibliographystyle{IEEEtran}
\bibliography{references}

\begin{thebibliography}{10}
\providecommand{\url}[1]{#1}
\csname url@samestyle\endcsname
\providecommand{\newblock}{\relax}
\providecommand{\bibinfo}[2]{#2}
\providecommand{\BIBentrySTDinterwordspacing}{\spaceskip=0pt\relax}
\providecommand{\BIBentryALTinterwordstretchfactor}{4}
\providecommand{\BIBentryALTinterwordspacing}{\spaceskip=\fontdimen2\font plus
\BIBentryALTinterwordstretchfactor\fontdimen3\font minus
  \fontdimen4\font\relax}
\providecommand{\BIBforeignlanguage}[2]{{%
\expandafter\ifx\csname l@#1\endcsname\relax
\typeout{** WARNING: IEEEtranM.bst: No hyphenation pattern has been}%
\typeout{** loaded for the language `#1'. Using the pattern for}%
\typeout{** the default language instead.}%
\else
\language=\csname l@#1\endcsname
\fi
#2}}
\providecommand{\BIBdecl}{\relax}
\BIBdecl

\bibitem{2023WWW_traffic}
K.-T. Foerster, T.~Marette, S.~Neumann, C.~Plant, Y.~Sadikaj, S.~Schmid, and
  Y.~Velaj, ``Analyzing the communication clusters in datacenters,'' in
  \emph{Proceedings of the ACM Web Conference 2023}, ser. WWW '23.\hskip 1em
  plus 0.5em minus 0.4em\relax New York, NY, USA: Association for Computing
  Machinery, 2023, p. 3022–3032.
\bibitem{2022Jupiter}
L.~e.~a. Poutievski, ``Jupiter evolving: transforming google's datacenter
  network via optical circuit switches and software-defined networking,'' in
  \emph{Proceedings of the ACM SIGCOMM 2022 Conference}, ser. SIGCOMM
  '22.\hskip 1em plus 0.5em minus 0.4em\relax New York, NY, USA: Association
  for Computing Machinery, 2022, p. 66–85.
\bibitem{singh2015jupiter}
A.~Singh, J.~Ong, A.~Agarwal, G.~Anderson, A.~Armistead, R.~Bannon, S.~Boving,
  G.~Desai, B.~Felderman, P.~Germano \emph{et~al.}, ``Jupiter rising: A decade
  of clos topologies and centralized control in google's datacenter network,''
  \emph{Proc. ACM SIGCOMM Computer Communication Review (CCR)}, vol.~45, no.~4,
  pp. 183--197, 2015.
\bibitem{f10}
V.~Liu, D.~Halperin, A.~Krishnamurthy, and T.~Anderson, ``F10: A fault-tolerant
  engineered network,'' in \emph{Proceedings of the 10th USENIX Conference on
  Networked Systems Design and Implementation}, ser. NSDI'13.\hskip 1em plus
  0.5em minus 0.4em\relax Berkeley, CA, USA: USENIX Association, 2013, pp.
  399--412.
\bibitem{bcube}
C.~Guo, G.~Lu, D.~Li, H.~Wu, X.~Zhang, Y.~Shi, C.~Tian, Y.~Zhang, and S.~Lu,
  ``{BCube: a high performance, server-centric network architecture for modular
  data centers},'' \emph{ACM SIGCOMM Computer Communication Review (CCR)},
  vol.~39, no.~4, pp. 63--74, 2009.
\bibitem{mdcube}
H.~Wu, G.~Lu, D.~Li, C.~Guo, and Y.~Zhang, ``{MDCube: a high performance
  network structure for modular data center interconnection},'' in \emph{ACM
  International Conference on Emerging Networking Experiments and Technologies
  (CONEXT)}, 2009, pp. 25--36.
\bibitem{xpander}
S.~Kassing, A.~Valadarsky, G.~Shahaf, M.~Schapira, and A.~Singla, ``Beyond
  fat-trees without antennae, mirrors, and disco-balls,'' in \emph{Proc. ACM
  SIGCOMM}, 2017, pp. 281--294.
\bibitem{jellyfish}
A.~Singla, C.-Y. Hong, L.~Popa, and P.~B. Godfrey, ``Jellyfish: Networking data
  centers, randomly,'' in \emph{Proc. USENIX Symposium on Networked Systems
  Design and Implementation (NSDI)}, vol.~12, 2012, pp. 17--17.
\bibitem{trace-complexity}
C.~Avin, M.~Ghobadi, C.~Griner, and S.~Schmid, ``On the complexity of traffic
  traces and implications,'' in \emph{Proc. ACM SIGMETRICS}, 2020.
\bibitem{2025CACM}
\BIBentryALTinterwordspacing\relax C.~Avin and S.~Schmid, ``Revolutionizing
  datacenter networks via reconfigurable topologies,'' \emph{Commun. ACM},
  vol.~68, no.~6, p. 44–53, Jun. 2025. [Online]. Available:
  \url{https://doi.org/10.1145/3708980}
\BIBentrySTDinterwordspacing
\bibitem{2024RDCN}
J.~Li, H.~Gong, F.~De~Marchi, A.~Gong, Y.~Lei, W.~Bai, and Y.~Xia,
  ``Uniform-cost multi-path routing for reconfigurable data center networks,''
  in \emph{Proceedings of the ACM SIGCOMM 2024 Conference}, ser. ACM SIGCOMM
  '24.\hskip 1em plus 0.5em minus 0.4em\relax New York, NY, USA: Association
  for Computing Machinery, 2024, p. 433–448.
\bibitem{2021SurveyRONs}
M.~Hall, K.-T. Foerster, S.~Schmid, and R.~Durairajan,
  ``\BIBforeignlanguage{English}{A survey of reconfigurable optical
  networks},'' \emph{\BIBforeignlanguage{English}{Optical Switching and
  Networking}}, vol.~41, Sep. 2021, publisher Copyright: {\textcopyright} 2021
  Elsevier B.V.
\bibitem{2024NegotiaToR}
C.~Liang, X.~Song, J.~Cheng, M.~Wang, Y.~Liu, Z.~Liu, S.~Zhao, and Y.~Cui,
  ``{NegotiaToR: Towards A Simple Yet Effective On-demand Reconfigurable
  Datacenter Network},'' in \emph{ACM SIGCOMM Conference}.\hskip 1em plus 0.5em
  minus 0.4em\relax New York, NY, USA: Association for Computing Machinery,
  2024, p. 415–432.
\bibitem{sigmetrics23mars}
V.~Addanki, C.~Avin, and S.~Schmid, ``Mars: Near-optimal throughput with
  shallow buffers in reconfigurable datacenter networks,'' in \emph{{ACM
  SIGMETRICS}}, 2023.
\bibitem{sigmetrics23duo}
J.~Zerwas, C.~Gyorgyi, A.~Blenk, S.~Schmid, and C.~Avin, ``Duo: A
  high-throughput reconfigurable datacenter network using local routing and
  control,'' in \emph{{ACM SIGMETRICS}}, 2023.
\bibitem{sirius}
H.~Ballani, P.~Costa, R.~Behrendt, D.~Cletheroe, I.~Haller, K.~Jozwik,
  F.~Karinou, S.~Lange, K.~Shi, B.~Thomsen \emph{et~al.}, ``Sirius: A flat
  datacenter network with nanosecond optical switching,'' in \emph{Proc. ACM
  SIGCOMM}, 2020, pp. 782--797.
\bibitem{projector}
M.~Ghobadi, R.~Mahajan, A.~Phanishayee, N.~Devanur, J.~Kulkarni, G.~Ranade,
  P.-A. Blanche, H.~Rastegarfar, M.~Glick, and D.~Kilper, ``{ProjecToR: Agile
  Reconfigurable Data Center Interconnect},'' in \emph{Proc. ACM
  SIGCOMM}.\hskip 1em plus 0.5em minus 0.4em\relax New York, NY, USA: ACM,
  2016, pp. 216--229.
\bibitem{2017WSS}
A.~Shakeri, M.~Garrich, A.~Bravalheri, D.~Careglio, J.~Sole-Pareta, and
  A.~Fumagalli, ``Traffic allocation strategies in wss-based dynamic optical
  networks,'' \emph{Journal of Optical Communications and Networking}, vol.~9,
  no.~4, pp. B112--B123, 2017.
\bibitem{2021nature}
A.~S. Raja, S.~Lange, M.~Karpov, K.~Shi, X.~Fu, R.~Behrendt, D.~Cletheroe,
  A.~Lukashchuk, I.~Haller, F.~Karinou, B.~Thomsen, K.~Jozwik, J.~Liu,
  P.~Costa, T.~J. Kippenberg, and H.~Ballani, ``Ultrafast optical circuit
  switching for data centers using integrated soliton microcombs,''
  \emph{Nature Communications}, vol.~12, no.~1, 2021.
\bibitem{infocom22lazy}
E.~Feder, I.~Rathod, P.~Shyamsukha, R.~Sama, V.~Aksenov, I.~Salem, and
  S.~Schmid, ``Lazy self-adjusting bounded-degree networks for the matching
  model,'' in \emph{Proc. IEEE Conference on Computer Communications
  (INFOCOM)}, 2022.
\bibitem{spaa21lazy}
------, ``Toward self-adjusting networks for the matching model,'' in
  \emph{33rd ACM Symposium on Parallelism in Algorithms and Architectures
  (SPAA)}, 2021.
\bibitem{avin2020online}
C.~Avin, C.~Griner, I.~Salem, and S.~Schmid, ``An online matching model for
  self-adjusting tor-to-tor networks,'' 2020.
\bibitem{infocom23matching}
K.~Hanauer, M.~Henzinger, L.~Ost, and S.~Schmid, ``Dynamic demand-aware link
  scheduling for reconfigurable datacenters,'' in \emph{{IEEE INFOCOM}},
  2023.
\bibitem{Hanauer_2022}
K.~Hanauer, M.~Henzinger, S.~Schmid, and J.~Trummer, ``Fast and heavy disjoint
  weighted matchings for demand-aware datacenter topologies,'' in \emph{IEEE
  INFOCOM 2022 - IEEE Conference on Computer Communications}.\hskip 1em plus
  0.5em minus 0.4em\relax IEEE, May 2022, p. 1649–1658.
\bibitem{2023TNSM}
\BIBentryALTinterwordspacing\relax T.~Mano, T.~Inoue, K.~Mizutani, and
  O.~Akashi, ``Redesigning the nonblocking clos network to increase its
  capacity,'' \emph{IEEE Trans. on Netw. and Serv. Manag.}, vol.~20, no.~3, p.
  2558–2574, Sep. 2023. [Online]. Available:
  \url{https://doi.org/10.1109/TNSM.2022.3228909}
\BIBentrySTDinterwordspacing
\bibitem{2025TNSM3}
E.~Oki, R.~Taniguchi, K.~Anazawa, and T.~Inoue, ``{Design of Multiple-Plane
  Twisted and Folded Clos Network Guaranteeing Admissible Blocking
  Probability},'' \emph{IEEE Transactions on Network and Service Management},
  vol.~22, no.~3, pp. 2278--2294, 2025.
\bibitem{ccr18san}
C.~Avin and S.~Schmid, ``Toward demand-aware networking: A theory for
  self-adjusting networks,'' in \emph{ACM SIGCOMM Computer Communication Review
  (CCR)}, 2018.
\bibitem{apocs21renets}
------, ``Renets: Statically-optimal demand-aware networks,'' in \emph{Proc.
  SIAM Symposium on Algorithmic Principles of Computer Systems (APOCS)},
  2021.
\bibitem{splaynet}
S.~Schmid, C.~Avin, C.~Scheideler, M.~Borokhovich, B.~Haeupler, and Z.~Lotker,
  ``Splaynet: Towards locally self-adjusting networks,'' \emph{IEEE/ACM Trans.
  Netw.}, vol.~24, no.~3, pp. 1421--1433, 2016.
\bibitem{2021TCCDisplaynet}
B.~Peres, O.~A. d.~O. Souza, O.~Goussevskaya, C.~Avin, and S.~Schmid,
  ``Distributed self-adjusting tree networks,'' \emph{IEEE Transactions on
  Cloud Computing}, 2021.
\bibitem{displaynet}
B.~Peres, O.~Souza, O.~Goussevskaia, S.~Schmid, and C.~Avin, ``Distributed
  self-adjusting tree networks,'' in \emph{Proc. IEEE INFOCOM}, 2019.
\bibitem{2021CBNet}
O.~A. d.~O. Souza, O.~Goussevskaia, and S.~Schmid, ``{CBNet: Minimizing
  Adjustments in Concurrent Demand-Aware Tree Networks},'' in \emph{IEEE Int.
  Parallel and Distributed Processing Sym. (IPDPS)}, 2021, pp. 382--391.
\bibitem{2022COMNETCBNet}
O.~A. de~O.~Souza, O.~Goussevskaia, and S.~Schmid, ``{CBNet: Demand-aware tree
  topologies for Reconfigurable Datacenter Networks},'' \emph{Computer
  Networks}, vol. 213, p. 109090, 2022.
\bibitem{disc17}
C.~Avin, K.~Mondal, and S.~Schmid, ``Demand-aware network designs of bounded
  degree,'' in \emph{Proc. International Symposium on Distributed Computing
  (DISC)}, 2017.
\bibitem{2023infocomOpticnet}
C.~A. Caldeira, O.~A. de~O.~Souza, O.~Goussevskaia, and S.~Schmid, ``{OpticNet:
  Self-Adjusting Networks for ToR-Matching-ToR Optical Switching
  Architectures},'' in \emph{IEEE INFOCOM 2023 - IEEE Conference on Computer
  Communications}, 2023, pp. 1--10.
\bibitem{2024tccOpticnet}
C.~Caldeira, O.~A. de~Oliveira~Souza, O.~Goussevskaia, and S.~Schmid, ``Optical
  self-adjusting data center networks in the scalable matching model,'' in
  \emph{IEEE/ACM Transactions on Cloud Computing (TCC)}, 2024.
\bibitem{splaytrees}
D.~Sleator and R.~Tarjan, ``Self-adjusting binary search trees,'' \emph{Journal
  of the ACM (JACM)}, vol.~32, no.~3, pp. 652--686, 1985.
\bibitem{cbtree}
Y.~Afek, H.~Kaplan, B.~Korenfeld, A.~Morrison, and R.~E. Tarjan, ``Cbtree: A
  practical concurrent self-adjusting search tree,'' in \emph{Proc.
  DISC}.\hskip 1em plus 0.5em minus 0.4em\relax Berlin, Heidelberg:
  Springer-Verlag, 2012, pp. 1--15.
\bibitem{pfabric}
\BIBentryALTinterwordspacing\relax M.~Alizadeh, S.~Yang, M.~Sharif, S.~Katti,
  N.~McKeown, B.~Prabhakar, and S.~Shenker, ``{PFabric: Minimal near-Optimal
  Datacenter Transport},'' in \emph{Proceedings of the ACM SIGCOMM 2013
  Conference on SIGCOMM}, ser. SIGCOMM ’13.\hskip 1em plus 0.5em minus
  0.4em\relax New York, NY, USA: Association for Computing Machinery, 2013, p.
  435–446. [Online]. Available:
  \url{https://doi.org/10.1145/2486001.2486031}
\BIBentrySTDinterwordspacing
\bibitem{Roy:2015}
A.~Roy, H.~Zeng, J.~Bagga, G.~Porter, and A.~C. Snoeren, ``Inside the social
  network's (datacenter) network,'' in \emph{Proceedings of the 2015 ACM
  SIGCOMM}, ser. SIGCOMM '15.\hskip 1em plus 0.5em minus 0.4em\relax New York,
  NY, USA: ACM, 2015, pp. 123--137.
\bibitem{rotornet}
W.~M. Mellette, R.~McGuinness, A.~Roy, A.~Forencich, G.~Papen, A.~C. Snoeren,
  and G.~Porter, ``Rotornet: A scalable, low-complexity, optical datacenter
  network,'' in \emph{Proc. ACM SIGCOMM}, 2017, pp. 267--280.
\bibitem{opera}
W.~M. Mellette, R.~Das, Y.~Guo, R.~McGuinness, A.~C. Snoeren, and G.~Porter,
  ``Expanding across time to deliver bandwidth efficiency and low latency,''
  \emph{arXiv preprint arXiv:1903.12307}, 2019.
\bibitem{2022STOC_AMIR}
D.~Amir, T.~Wilson, V.~Shrivastav, H.~Weatherspoon, R.~Kleinberg, and
  R.~Agarwal, ``Optimal oblivious reconfigurable networks,'' in \emph{{}Proc.
  of the 54th Annual ACM SIGACT Symposium on Theory of Computing}, 2022.
\bibitem{sigmetrics22cerberus}
C.~Griner, J.~Zerwas, A.~Blenk, S.~Schmid, M.~Ghobadi, and C.~Avin, ``Cerberus:
  The power of choices in datacenter topology design (a throughput
  perspective),'' in \emph{Proc. ACM SIGMETRICS}, 2022.
\bibitem{polatis}
``{Polatis Series $6000$},'' \url{www.polatis.com}, [Online; accessed
  10-May-2022].
\bibitem{2024TNSM}
K.~Aykurt, J.~Zerwas, A.~Blenk, and W.~Kellerer, ``{When TCP Meets
  Reconfigurations: A Comprehensive Measurement Study},'' \emph{IEEE
  Transactions on Network and Service Management}, vol.~21, no.~2, pp.
  1372--1386, 2024.
\bibitem{2022TNSM}
J.~Zerwas, I.~Poese, S.~Schmid, and A.~Blenk, ``{On the Benefits of Joint
  Optimization of Reconfigurable CDN-ISP Infrastructure},'' \emph{IEEE
  Transactions on Network and Service Management}, vol.~19, no.~1, pp.
  158--173, 2022.
\bibitem{2019SODA}
C.~Levy and R.~Tarjan, ``A new path from splay to dynamic optimality,'' in
  \emph{Proceedings of the 2019 Annual ACM-SIAM Symposium on Discrete
  Algorithms (SODA)}, pp. 1311--1330.
\bibitem{Tango_SICOMP}
E.~D. Demaine, D.~Harmon, J.~Iacono, and M.~P\v{a}tra\c{s}cu, ``Dynamic
  optimality---almost,'' \emph{SIAM Journal on Computing}, vol.~37, no.~1, pp.
  240--251, May 2007.
\bibitem{2025TNSM2}
X.~Wang, H.~Shen, and H.~Tian, ``{Optimal Partitioning of Traffic Demand for
  Coflow Scheduling in Hybrid Switches},'' \emph{IEEE Transactions on Network
  and Service Management}, vol.~22, no.~6, pp. 5972--5986, 2025.
\end{thebibliography}

\end{document}